\documentclass[12pt]{article}
\usepackage{mathrsfs}
\usepackage{appendix}
\usepackage{natbib}
\usepackage{extarrows}
\usepackage{graphicx,setspace,lscape,longtable,epstopdf,xr}
\usepackage{mathrsfs,amsmath,amsthm,amssymb,color}
\usepackage{epsfig,graphicx,pdfpages}
\usepackage{rotating}
\usepackage{float}
\usepackage{bm}
\usepackage{ragged2e}
\usepackage{todonotes}
\usepackage{hyperref}
\hypersetup{
colorlinks=true,
linkcolor=blue,
citecolor=blue,
urlcolor=blue}
\usepackage{booktabs}
\usepackage{algorithm}
\usepackage{algpseudocode}
\usepackage{amsmath}
\usepackage{graphicx}
\usepackage{subfigure}
\usepackage{hyperref}
\bibpunct{(}{)}{;}{a}{,}{,}

\newtheorem{theorem}{Theorem}
\newtheorem{lemma}{Lemma}
\newtheorem{proposition}{Proposition}

\newcommand{\csection}[1]
{\begin{center}
\stepcounter{section}
{\bf\large\arabic{section}. #1}
\end{center}
}

\newcommand{\scsection}[1]
{\begin{center}
{\bf\large #1}
\end{center}
}

\newcommand{\csubsection}[1]{
\begin{center}
\stepcounter{subsection}
{\it\arabic{section}.\arabic{subsection}. #1}
\end{center}
}

\newcommand{\scsubsection}[1]{
\begin{center}
\stepcounter{subsection}
{\it #1}
\end{center}
}
\def\one{{\bf 1}}
\def\bA{{\mathcal A}}

\def\beq{\begin{equation}}
\def\eeq{\end{equation}}
\def\beqr{\begin{eqnarray}}
\def\eeqr{\end{eqnarray}}
\def\beqrs{\begin{eqnarray*}}
\def\eeqrs{\end{eqnarray*}}
\def\bet{\begin{theorem}}
\def\eet{\end{theorem}}
\def\bel{\begin{lemma}}
\def\eel{\end{lemma}}
\def\bep{\begin{proposition}}
\def\eep{\end{proposition}}
\def\bg{\begin{figure}[tbph]\begin{center}}
\def\eg{\end{center}\end{figure}}

\def\bc{\begin{center}}
\def\ec{\end{center}}

\def\e{\mathbf{e}}

\newtheorem{remark}{Remark}
\newtheorem{corollary}{Corollary}

\def\wt{\widetilde}

\def\wh{\widehat}

\def\ol{\overline }

\def\mC{\mathcal C}

\def\mP{\mathbb P}
\def\mD{\mathcal D}

\def\mR{\mathbb{R}}
\def\cP{\mathcal{P}}

\def\mL{\mathcal L}
\def\mS{\mathbb S}

\def\mM{\mathcal M}

\def\mF{\mathcal F}
\def\cR{\mathcal{R}}

\def\mS{\mathcal S}

\def\diag{\mbox{diag}}

\newcommand{\RNum}[1]{\uppercase\expandafter{\romannumeral #1\relax}}

\def\bg{\mbox{\boldmath $g$}}

\def\zero{\mathbf{0}}
\def\defeq{\stackrel{\mathrm{def}}{=}}  

\numberwithin{equation}{section}

\begin{document}
\begin{center}
{\bf\Large Distributed Community Detection for Large Scale Networks Using Stochastic Block Model}\\
\bigskip
Shihao Wu$^1$, Zhe Li$^{1}$, and Xuening Zhu$^1$

{\it $^1$School of Data Science, Fudan University, Shanghai, China}

{\it\small

}

\end{center}

\begin{footnotetext}[1] {Shihao Wu and Zhe Li are joint first authors.
Xuening Zhu is corresponding author ({\it xueningzhu@fudan.edu.cn}).
Xuening Zhu is supported by the National Natural Science Foundation of China (nos. 11901105, 71991472, U1811461), the Shanghai Sailing Program for Youth Science and Technology Excellence (19YF1402700), and the Fudan-Xinzailing Joint Research Centre for Big Data, School of Data Science, Fudan University.}
\end{footnotetext}

\begin{singlespace}
\begin{abstract}

With rapid developments of information and technology, large scale network data are ubiquitous.
In this work we develop a distributed spectral clustering algorithm for community detection in large scale networks.
To handle the problem, we distribute $l$ pilot network nodes on the master server and the others on worker servers.
A spectral clustering algorithm is first conducted on the master to select pseudo centers.
The indexes of the pseudo centers are then broadcasted to workers to complete distributed community detection task
using a SVD type algorithm.
The proposed distributed algorithm has three merits.
First, the communication cost is low since only the indexes of pseudo centers are communicated.
Second, no further iteration algorithm is needed on workers and hence it does not suffer from problems as initialization and non-robustness.
Third, both the computational complexity and the storage requirements are much lower compared to using the whole adjacency matrix.
A Python package \textsf{DCD} (\href{https://github.com/Ikerlz/dcd}{www.github.com/Ikerlz/dcd}) is developed to implement the distributed algorithm for a Spark system.
Theoretical properties are provided with respect to the estimation accuracy and mis-clustering rates.
Lastly, the advantages of the proposed methodology are illustrated by experiments on a variety of synthetic and empirical datasets.\\


  \noindent {\bf KEY WORDS: } Large scale network; Community detection; Distributed spectral clustering; Stochastic block model; Distributed system.\\

\end{abstract}
\end{singlespace}

\newpage

\csection{INTRODUCTION}

Large scale networks have become more and more popular in today's world.
Recently, network data analysis receives great attention in a wide range of applications,
which include but not limited to social network analysis \citep{sojourner2013identification,liu2017peer,zhu2020multivariate},
biological study \citep{marbach2010revealing,marbach2012wisdom},
financial risk management \citep{hardle2016tenet,zou2017covariance} and many others.

Among the existing literature for large scale network data,
the stochastic block model (SBM) is widely used due to its simple form and great usefulness \citep{holland1983stochastic}.
In a SBM, the network nodes are partitioned into $K$ communities according to their connections.
Within the same community, nodes are more likely to form edges with each other.
On the other hand, the nodes from different communities are less likely to form connections.
Understanding the community structure is vital in a variety of fields.
For instance, in social network analysis, users from the same community are likely to share similar social interests.
As a consequence, particular marketing strategies can be applied based on their community memberships.

Statistically, the communities in the SBM are latent hence need to be detected.
One of the most fundamental problems in the SBM is to recover community memberships from the observed network relationships.
To address this issue, researchers have proposed various estimation methods to accomplish this task.
For instance, \cite{zhao2012consistency}, \cite{amini2013pseudo} and  \cite{bickel2009nonparametric} adopted likelihood based methods and proved asymptotic properties.
Other approaches include convex optimization \citep{chen2012clustering},
methods of moments \citep{anandkumar2014tensor},
spectral clustering \citep{lei2015consistency,jin2015fast,lei2020consistency} and many others.

Among the approaches, spectral clustering \citep{von2007tutorial,balakrishnan2011noise,rohe2011spectral,lei2015consistency,jin2015fast,sarkar2015role,lei2020consistency} is one of the most widely used methods for community detection.
Particularly, it first performs eigen-decomposition using the adjacency matrix or the graph Laplacian matrix.
Then the community memberships are estimated by further applying a $k$-means algorithm to the first several leading eigenvectors.
Theoretically, both \cite{rohe2011spectral} and \cite{lei2015consistency} have studied the consistency of spectral clustering under stochastic block models.

Despite the usefulness of spectral clustering on community detection problem, the procedure is computationally demanding especially when the network is of large scale.
In the meanwhile, with rapid developments of information and technology, large scale network data
are ubiquitous.
On one hand, handling such enormous datasets requires great computational power and storage capacity.
Hence, it is nearly impossible to complete statistical modelling tasks on a central server.
On the other hand, the concerns of privacy and ownership issues require the datasets to be distributed
across different data centers.
In the meanwhile, due to the distributed storage of the datasets, constraints on communication budgets
also post great challenges on statistical modelling tasks.
Therefore, developing distributed statistical modelling methods which are efficient with low computation and communication cost is important.

In recent literature, a surge of researches have emerged to solve the distributed statistical modelling problems.
For instance, to conduct distributed regression analysis, both one-shot and iterative distributed algorithms are
designed and studied
\citep{zhang2013communication,liu2014distributed,chang2017divide,chang2017distributed}.
 Recently, a number of works start to pay attention to
distributed community detection tasks.
For example, \citep{yang2015divide}
proposed a divide and conquer framework for distributed graph clustering.
The algorithm first partitions the nodes into $m$ groups and conduct clustering on each subgraph in parallel.
Then, they merge the clustering result using a trace-norm based optimization approach.
\cite{mukherjee2017two} devised two algorithms to solve the problem, which first sample several subgraphs and conduct clustering, and then patch the results into a single clustering result.
Other methods for scalable community detection tasks include: random walk based algorithms \citep{fathi2019efficient},
randomized sketching methods \citep{rahmani2020scalable} and so on.


In this work, we propose a distributed community detection (DCD) algorithm.
The distributed system typically consists of a master server and multiple worker servers.
In each round of computation, the master server is responsible to broadcast tasks to workers,
then the workers conduct computational tasks using local datasets and communicate the results to the master.
More specifically, we distribute the network nodes together with their
network relationships on both masters and workers.
Specifically, on the master server we distribute $l$ network nodes, who are referred to as {\it pilot nodes.}
The network relationships among the pilot nodes are stored on the master server.
On the $m$th worker, we distribute $n_m$ network nodes together with $l$ pilot nodes.
The network relationships between the $n_m$ network nodes and the pilot nodes are recorded.
Compared with storing the whole network relationships, we resort to storing only a partial network, which leads to much lower storage requirements.

\begin{figure}[H]
	\centering
	\includegraphics[width=1\textwidth]{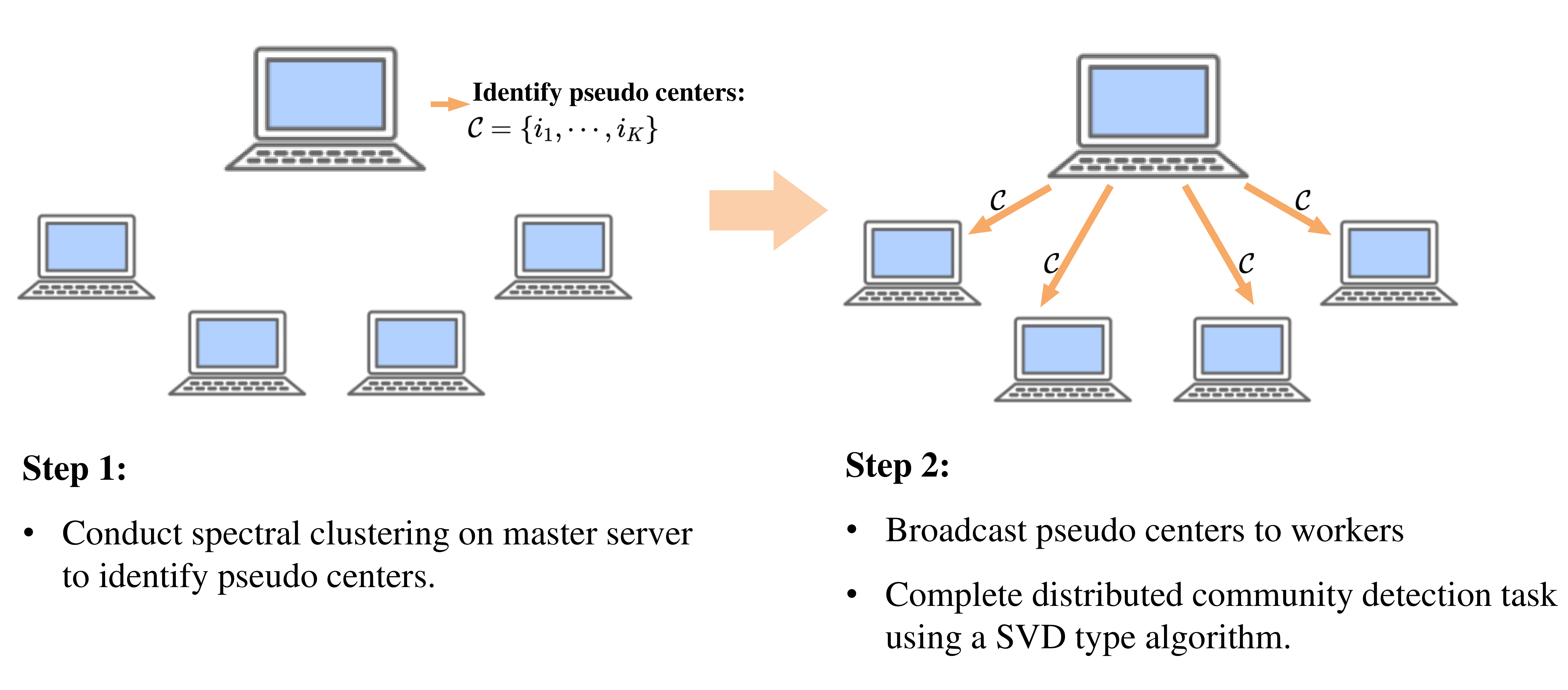}
	\caption{\small Illustration of distributed community detection algorithm.
One time communication is required between master and workers.
	}\label{fig_broadcast}
\end{figure}

The distributed community detection is then conducted as follows.
First, we perform a spectral clustering on the master server using $l$ pilot nodes.
During this step, $K$ {\it pseudo centers} are identified.
The pseudo centers are identified as the pilot nodes most close to the clustering centers.
Next, we broadcast the indexes of the pseudo centers to workers to further complete the community detection task
by using a SVD type algorithm.
The basic steps of the algorithm are summarized in Figure \ref{fig_broadcast}.
Compared with the existing approaches,
our algorithm has the following three merits.
First, the communication cost is low since only the indexes of pseudo centers are communicated and only one time communication is used.
Second, no further iteration algorithm is needed on workers and hence it does not suffer from problems as initialization and non-robustness.
Third, the total computational complexity is of order $O(Ml^3+\sum_{m=1}^{M}n_ml^2)$,
 where $M$ is the number of workers.
 Therefore the computational cost is low as long as
the size of pilot nodes is well controlled.
We would like to remark that the proposed algorithm can be applied not only on distributed systems,
 but also on a single computer with memory constraint.
Theoretically, we establish upper bounds of (a) the singular vector estimation error and (b) the number of mis-clustering nodes.
Extensive numerical study and comparisons are presented to illustrate the computational power of the proposed methodology.

The article is organized as follows. In Section 2, we introduce stochastic block model and our
distributed community detection algorithm.
In Section 3, we develop the theoretical properties of the estimation accuracies of the community detection task.
In Section 4 and 5, we study the performance of our algorithm via simulation
and real data analysis.
Section 6 concludes the article with a discussion. All proofs and technique lemmas are relegated to the Appendix.

\csection{DISTRIBUTED SPECTRAL CLUSTERING FOR STOCHASTIC BLOCK MODEL}

\csubsection{Stochastic Block Model and Spectral Clustering}

Consider a large scale network with $N$ nodes, which can be clustered into $K$ communities.
For each node $i$, let $g_i\in \{1,\cdots, K\}$ be its community label.
A stochastic block model is parameterized by a {\it membership matrix}
$\Theta = (\Theta_1,\cdots, \Theta_N)^\top\in\mR^{N\times K}$ and a {\it connectivity matrix} $B \in\mR^{K\times K}$ (with full rank).
For the $i$th row of $\Theta$,
only the $g_i$th element takes 1 and the others are 0.
In addition, the connectivity matrix $B$ characterizes the connection probability between communities.
Specifically, the connection probability between the $k$th and $l$th community is  $B_{kl}$.
The edge $A_{ij}$ between the node $i$ and $j$ is generated independently from
$\mbox{Bernoulli}(B_{g_ig_j})$ distribution.
The adjacency matrix is then defined as $A = (A_{ij})$.
By using the adjacency matrix, the Laplacian matrix $L$ can be defined as
$L = D^{-1/2}AD^{-1/2}$, where $D$ is a diagonal matrix with the $i$th diagonal element
being $D_{ii} = \sum_{j} A_{ij}$.

Define $\mathcal{A}=\mathbb{E}(A)$ and $\mD=\mathbb{E}(D)$ as the population leveled counterparts of $A$ and $D$.
Accordingly let $\mL=\mD^{-1/2}\mathcal{A}\mD^{-1/2}$. For a matrix $X\in\mR^{m\times n}$, denote $X_i\in \mR^n$ as the $i$th row of matrix $X$.
The following Lemma shows the connection between the membership matrix and
the eigenvector matrix of $\mL$.

\begin{lemma}\label{lemma_SBM_Spectrum}
The eigen-decomposition of $\mL$ takes the form $\mL=U\Lambda U^{\top}$, where $U=(U_1,\cdots, U_N)^\top\in\mR^{N\times K}$ collects the eigen-vectors and $\Lambda\in \mR^{K\times K}$ is a diagonal matrix.
Further we have
	$U=\Theta \mu$,
	where $\mu$ is a $K\times K$ orthogonal matrix and
$\Theta_i = \Theta_j$ if and only if $U_i = U_j$.
\end{lemma}

The proof of Lemma \ref{lemma_SBM_Spectrum} is given by  \cite{rohe2011spectral}.
By Lemma \ref{lemma_SBM_Spectrum}, it can be concluded that
$U$ only has $K$ distinct rows and the $i$th row is equal to the $j$th row if the corresponding two nodes belong to the same community.
Accordingly, let $\wh U \in\mR^{N\times K}$ denote the $K$ eigenvectors of $L$
with top $K$ absolute eigenvalues.
Under mild conditions, one can show that $\wh U$ is a slightly perturbed version of $U$ and thus has roughly $K$ distinct rows as well.
Applying a $k$-means clustering algorithm to $\wh U$, we are then able to estimate the membership matrix.
The spectral clustering algorithm is summarized in Algorithm \ref{sc_all}.

\begin{center}
\begin{minipage}{13.5cm}
	  \begin{algorithm}[H]
  \caption{Spectral Clustering for SBM}\label{sc_all}
  \begin{algorithmic}[1]
    \Require
      Adjacency matrix $A$;
      number of communities $K$;
      approximation error $\varepsilon$.
    \Ensure
      Membership matrix $\widehat{\Theta}$.
    \State Compute Laplacian matrix $L$ based on $A$.
    \State Conduct eigen-decomposition of $L$ and extract the top $K$ eigenvectors (i.e., $\wh U$).
    \State
    Conduct $k$-means algorithm using $\wh U$ and then output the estimated membership matrix $\wh \Theta$.
  \end{algorithmic}
\end{algorithm}
\end{minipage}
\end{center}

Despite the usefulness, the classical spectral clustering method for the SBM is computationally intensive
with computational complexity in the order $O(N^3)$.
Hence it is hard to apply in the large scale networks.
In the following we aim to develop a distributed spectral clustering algorithm
for the SBM model.
Specifically, we first introduce a pilot network spectral clustering algorithm on the master server
in Section 2.2.
Then we elaborate the communication mechanism and computation on workers for the distributed community detection task in Section 2.3 and Section 2.4.

\csubsection{Pilot Network Spectral Clustering on Master Server}

For the distributed community detection task,
we first conduct a pilot-based spectral clustering on the master server.
{Suppose we have $l$ network nodes on the master,
which are referred to as {\it pilot nodes}.
In addition we distribute the pilot nodes both on master and workers.
 In the distributed system, the adjacency matrix is distributed as in Figure \ref{adj_nodes}.
As a result, compared to storing the whole network relationships, only a sub-adjacency matrix (i.e., partial network) is stored.
This leads to a much lower storage requirement.

\begin{figure}[h]
  \centering
  \includegraphics[width=1\textwidth]{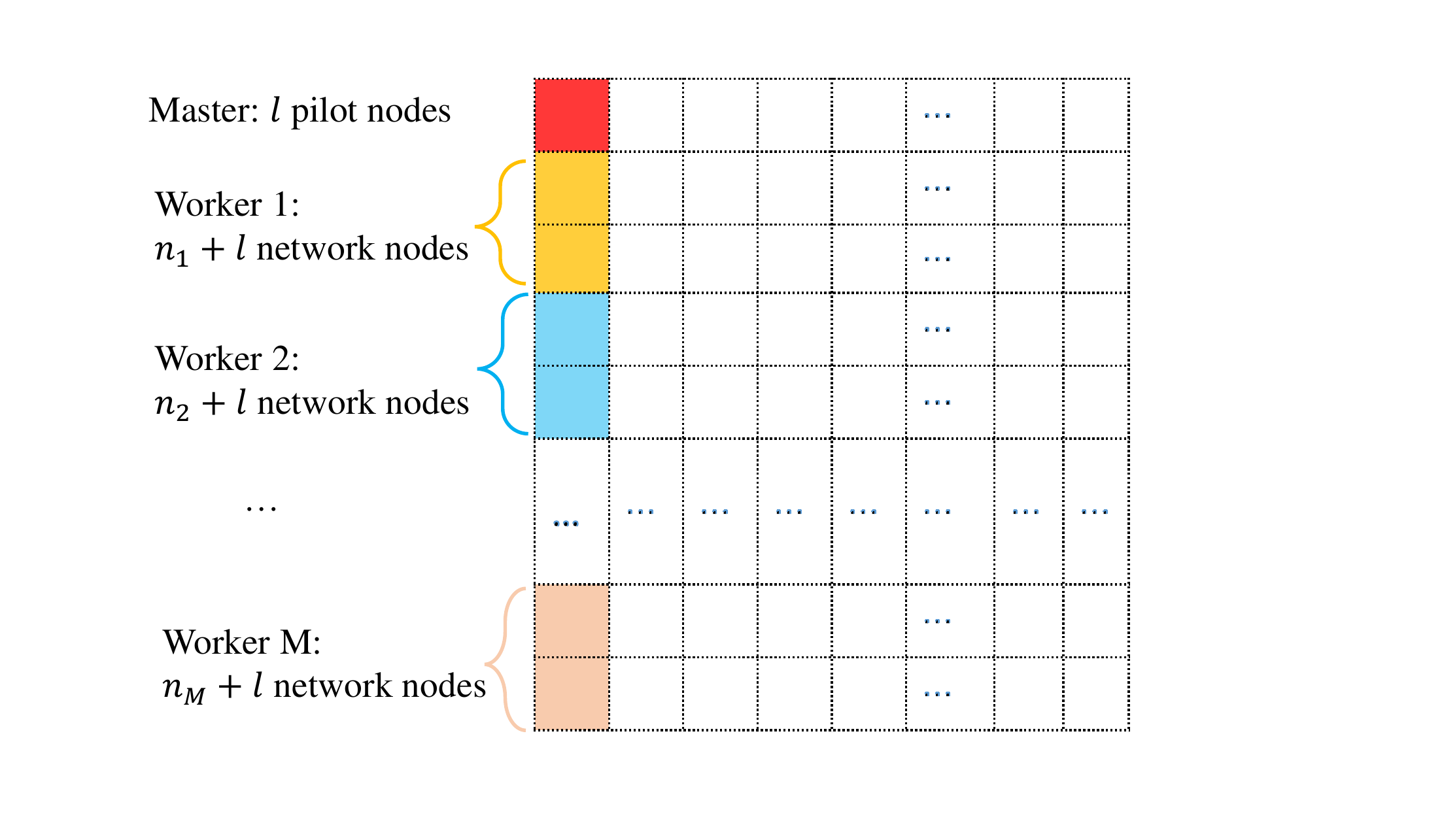}
  \caption{\small Distributed adjacency matrix $A$ in the distributed system. On master server, $l$ pilot nodes are distributed. On the $m$th worker, the network relationships between the $n_m+l$ network nodes and pilot nodes are stored.
}\label{adj_nodes}
\end{figure}

Let $m_k  = \sum_{i=1}^N I(g_i = k)$ be the number of nodes in the $k$th community.
In addition, define $n_{0k}$ be the number of pilot nodes in the $k$th community.
Without loss of generality, we assume $n_{0k}/m_k = r_0$ for $k = 1,\cdots, K$.
Consequently, the relative size of each community (i.e. distribution of memberships) is the same for the pilot nodes.

Subsequently, define the adjacency matrix among the pilot nodes as $A_0\in \mR^{l\times l}$.
Denote the corresponding Laplacian matrix $L_0$ as
$L_0=D_0^{-1/2}A_0D_0^{-1/2}$,
where $D_0 = \diag\{D_{0,11},\cdots, D_{0,ll}\}$ with
$D_{0,ii} = \sum_j A_{0,ij}$.
Accordingly, define $\mD_0 = E(D_0)$, $\bA_0 = E(A_0)$,
and $\mL_0 = \mD_0^{-1/2} \bA_0 \mD_0^{-1/2}$.

%

By Lemma 3.1 in \cite{rohe2011spectral}, the eigen-decomposition of $\mL_0$ takes the form as
$\mL_0 = U_0 \Lambda_0U_0^{\top}$, where
$\Lambda_0\in \mR^{K\times K}$ and
$U_0\in \mR^{l\times K}$ has
$K$ distinct rows.
We collect the $K$ distinct rows in the matrix $U_0^{(K)}\in\mR^{K\times K}$.
In addition, let $U^{(K)}$ collect $K$ distinct rows of $U$.
The following proposition establishes the relationship between $U_0^{(K)}$ and  $U^{(K)}$.

\bep\label{prop_U}
Under the assumption that $n_{0k}/m_k = r_0$ for $k = 1,\cdots, K$, we have $U_0^{(K)} = r_0^{-1/2} U^{(K)}$.
\eep
The proof of Proposition \ref{prop_U} is given in Appendix B.1.
By Proposition \ref{prop_U}, it could be concluded that $U_0^{(K)}$ is equivalent
to $U^{(K)}$ up to a ratio $r_0^{-1/2}$.
Empirically, by conducting a spectral clustering algorithm on the master,
we are able to cluster the pilot nodes correctly with a high probability \citep{rohe2011spectral}.

%
%

\csubsection{Pseudo Centers and Communication Mechanism}

After clustering pilot nodes on the master, we then broadcast the clustering results to workers to further complete community detection on workers.
To conduct the task, $K$ {\it pseudo centers} are selected for broadcasting as in Step 2 of Algorithm \ref{distribute_SBM}.
To be more specific, define the clustering centers of $\wh U_0$ (eigenvector matrix of $L_0$) after the $k$-means algorithm as $\wh C_0 = (\wh C_{0k}: 1\le k \le K)^\top$.
Then the index of the $k$th pseudo center is defined by
$i_k = \arg\min_i \|\wh U_{0i} - \wh C_{0k}\|$,
which is the closest node to the center of the $k$th cluster.
The pseudo centers are {\it pseudo} in the sense that they are not exactly the clustering centers
but the closest nodes to the centers.
As a result, they could be treated as the most representative nodes for each community. The indexes of pseudo centers are recorded as $\mC = \{i_1,\cdots, i_K\}$.

\begin{remark}

  Note that in the communication step, we only broadcast the indexes of pseudo centers instead of
clustering centers $\wh C_0$.
There are two advantages for doing so.
First, the communication cost is low compared to broadcasting $\wh C_0$.
Specifically, only $K$ integers need to be communicated.
Second, even though the clustering center matrix $\wh C_0$ is broadcasted, we still need to know a rotation matrix for further clustering on workers.
Instead, by broadcasting pseudo centers, we no longer need to estimate the rotations
but only use the pseudo center indexes on workers.
The detailed procedure is presented in the next section.
\end{remark}

\csubsection{Community Detection on Workers}

Suppose we distribute $n_m$ network nodes as well as the pilot nodes on the $m$th worker.
Let $\mathcal{P}$ collect the indexes of pilot nodes and $\mM_m$ collect the indexes of $n_m$ network nodes on the $m$th worker.
Denote $\mS_m = \cP\cup \mM_m$
with $|\mS_m| = \ol n_m$,
where $\ol n_m = l+n_m$.
Particularly on the $m$th worker, we store the network relationships between nodes in $\mS_m$
and the pilot nodes in $\cP$.
Denote the corresponding sub-adjacency matrix as $A^{(\mS_m)} \in \mR^{\ol n_m\times l}$.
Without loss of generality, we permute the row indexes of $A^{(\mS_m)}$ to ensure
that $A^{(\mS_m)} = (A_1^{(\mS_m)\top},A_2^{(\mS_m)\top})^\top$ with $A_1^{(\mS_m)} = A_0$.
As a result, the first $l$ rows of $A^{(\mS_m)}$ (i.e., $A_1^{(\mS_m)}$) store the adjacency matrix for the pilot nodes, and the rest (i.e., $A_2^{(\mS_m)}$) records the network relationship between the other $n_m$ nodes (i.e., $\mM_m$)
and the $l$ pilot nodes.

Let $D_{ii}^{(\mS_m)} = \sum_j A_{ij}^{(\mS_m)}$
and $F_{jj}^{(\mS_m)} = \sum_i A_{ij}^{(\mS_m)}$
be the out- and in-degrees of node $i$ and $j$ in the subnetwork on worker $m$.
Correspondingly, define $D^{(\mS_m)} = \diag\{D_{ii}^{(\mS_m)}:1\le i\le \ol n_m\}\in \mR^{\ol n_m \times \ol n_m}$
and $F^{(\mS_m)} = \diag\{
F_{jj}^{(\mS_m)}: 1\le j \le l\}\in \mR^{l\times l}$.
Then a Laplacian version of $A^{(\mS_m)}$ is given by
$L^{(\mS_m)} = (D^{(\mS_m)})^{-1/2} A^{(\mS_m)} (F^{(\mS_m)})^{-1/2}
\in \mR^{\ol n_m \times l}$.

Given the Laplacian matrix, we further perform the clustering algorithm on workers.
First, we conduct a singular value decomposition (SVD) using $L^{(\mS_m)}$.}
Note that the SVD can be done very efficiently as follows.
First, we conduct an eigenvalue decomposition on $L^{(\mS_m)^\top}L^{(\mS_m)}\in\mR^{l\times l}$ with computational complexity in the order of $O(l^3)$.
This leads to $L^{(\mS_m)^\top}L^{(\mS_m)} = \wt V_m\wt \Lambda_m\wt V_m^\top$, where $\wt V_m\in\mR^{l\times l}$ is the
 right singular vectors of $L^{(\mS_m)}\in\mR^{l\times l}$ and $\wt \Lambda_m\in\mR^{l\times l}$ is a diagonal matrix.
Then the left singular vectors can be efficiently computed by
$ \wt U_m = L^{(\mS_m)}\wt V_m\wt \Lambda_m^{-1/2}$ with computational complexity $O(n_ml^2)$.
Next, let $\wh U^{(\mS_m)}\in \mR^{\ol n_m\times K}$
collect the
top $K$ left singular vectors in $\wt U_m$.
Then we assign each node to the cluster with the closest pseudo center.
Specifically, recall the indexes of the pseudo centers are collected by
$\mC = \{i_1,\cdots, i_K\}$.
As a result, for the $i$th ($l+1\le i\le \ol n_m$) node in $\mS_m$,
the cluster label $g_i$ is estimated by
\beq
\wh g_i = \arg\min_{1\le k\le K, i_k\in \mC}\big\|\wh U_{i}^{(\mS_m)} -\wh U_{i_{k}}^{(\mS_m)}\big \|^2.\label{est_label}
\eeq
An obvious merit of (\ref{est_label}) is that no further iteration algorithms (e.g., $k$-means)
are needed for clustering.
It makes the clustering results more stable and computationally efficient.
The procedure for community detection on workers is summarized in Step 3 of {Algorithm \ref{distribute_SBM}.}

\begin{remark}
We compare our algorithm with the divide-and-conquer algorithms proposed by \cite{yang2015divide} and \cite{mukherjee2017two} in terms of three aspects.
First, both of their algorithms are based on subgraph clustering and merging methods.
The local computational burden can be heavy and further tuning parameter procedures are involved.
In contrast, we conduct the spectral clustering only once on master and we do not  resort to further clustering procedures on workers.
As a result, we do not suffer from load unbalancing problems due to
asynchronous computing times on different workers as well as parameter tuning problems.
Second, both \cite{yang2015divide} and \cite{mukherjee2017two} require to communicate the clustering results of subgraphs to the master.
Hence the communication cost is at least $O(n_m)$, where $n_m$ is number of nodes in the $m$th subgraph.
On contrary, the communication cost of our algorithm is $O(K)$, which is much lower.
Lastly, both the methods of \cite{yang2015divide} and \cite{mukherjee2017two} require to store the whole network on the master, while we only need to store a partial network on the master.
As a consequence, our method is more friendly for privacy protection.
\end{remark}

\begin{center}
\begin{minipage}{13.5cm}
	  \begin{algorithm}[H]
  \caption{Distributed Community Detection (DCD) for SBM}\label{distribute_SBM}
  \begin{algorithmic}[1]
    \Require
      Adjacency matrix $A_0$;
      sub-adjacency matrices \{$A^{(\mS_m)}$\}$_{m=1,...,M}$;
      number of communities $K$;
      approximation error $\varepsilon$.
    \Ensure
      Membership matrix $ \wh\Theta$
    \begin{itemize}
        \item [{\sc Step 1}] {\sc Pilot-based Network Spectral Clustering on Master Server}
        \begin{itemize}
      \item [{\sc Step 1.1}] Conduct eigen-decomposition of $L_0$ and extract the top $K$ eigenvectors (denoted in matrix $\wh U_0$).
      \item [{\sc Step 1.2}] {Conduct $k$-means algorithm and obtain clustering centers $\wh C_0=\big(\wh C_{0 k}: 1 \leq k \leq K\big)^{\top}$.}
     \end{itemize}
     \end{itemize}
    \begin{itemize}
        \item [{\sc Step 2}] {\sc Broadcast Pseudo Centers to Workers}
        \begin{itemize}
      \item [{\sc Step 2.1}] Determine the indexes of the $k$th pseudo centers as
      $i_k = \arg\min_{i} \|\wh U_{0i} - \wh C_{0k}\|_2^2$.
      \item [{\sc Step 2.2}] Broadcast the index set of pseudo centers $\mC = \{i_1,\cdots, i_K\}$ to workers.
       \end{itemize}
    \end{itemize}
    \begin{itemize}
        \item [{\sc Step 3}] {\sc Community Detection on Workers}
        \begin{itemize}
      \item [{\sc Step 3.1}] Perform singular value decomposition using $L^{(\mS_m)}$
      and denote the top $K$ left singular vector matrix as $\wh U^{(\mS_m)}$.
      \item [{\sc Step 3.2}] Use (\ref{est_label}) to obtain the estimated community labels.
       \end{itemize}
    \end{itemize}

  \end{algorithmic}
\end{algorithm}
\end{minipage}
\end{center}

\csection{THEORETICAL PROPERTIES}

In this section, we discuss the accuracy of the clustering algorithm.
We first establish the theoretical properties of the procedure on the population level.
Next, the convergence of singular vectors is given, which is the key for establishing the consistent clustering result.
Lastly, we derive error bounds on the mis-clustering rates.

\csubsection{Theoretical Properties on Population Level}

To motivate the study, we first discuss the theoretical properties on the population level.
Define $\bA^{(\mS_m)} = E(A^{(\mS_m)})$,
$\mD^{(\mS_m)} = E(D^{(\mS_m)})\in \mR^{\ol n_m \times \ol n_m}$
and $\mF^{(\mS_m)} = E(F^{(\mS_m)})\in \mR^{l\times l}$.
In addition, the normalized population adjacency matrix is defined by
$\mL^{(\mS_m)} = (\mD^{(\mS_m)})^{-1/2} \bA^{(\mS_m)}(\mF^{(\mS_m)})^{-1/2}$.
Suppose the singular value decomposition of $\mL^{(\mS_m)}$ is
$\mL^{(\mS_m)} = U^{(\mS_m)}\Lambda^{(\mS_m)}(V^{(\mS_m)})^\top$,
where $U^{(\mS_{m})}\in \mR^{\ol n_m\times K}$ and
$V^{(\mS_{m})}\in \mR^{l\times K}$ are left and right eigenvectors respectively.
In the following proposition we show that $U^{(\mS_m)}$ has $K$ distinct rows and
could identify the memberships of the nodes uniquely.

\bep\label{prop_membership}
Let $\Theta^{(\mS_{m})}\in\mR^{\ol n_m\times K}$ be the membership matrix on the  $m$th worker.
Then we have
$U^{(\mS_{m})}=\Theta^{(\mS_{m})}\mu$,
where $\mu\in \mR^{K \times K}$ is a rotation matrix, and
\begin{equation}\nonumber
{\mu^{\top}\Theta^{(\mS_{m})}_i=\mu^{\top}\Theta^{(\mS_{m})}_j} \Leftrightarrow \Theta^{(\mS_{m})}_i=\Theta^{(\mS_{m})}_j.
\end{equation}
\eep
\noindent
Proof of Proposition \ref{prop_membership} is given in Appendix B.2.
Proposition \ref{prop_membership}
implies that the singular vectors could play the same role as the eigenvectors
of the adjacency matrix in the community detection.

We then build the connection between $U^{(\mS_m)}$ with the eigenvector matrix $U$ of $\mL$, i.e.,  $\mL = U \Lambda U^\top$.
Denote $U_m = (U_i: i\in \mS_m)^\top \in\mR^{\ol n_m\times K}$
as the submatrix of $U$ whose row indexes are in $\mS_m$.
The connection could be built between $U^{(\mS_m)}$ and $U_m$.
Denote $\ol n_{mk}$ as the number of nodes on the $m$th worker belonging to the $k$th community.
If we have $\ol n_{mk}/m_k$s are equivalent over $1\le k \le K$.
Then it could be easily verified as Proposition \ref{prop_U} that
$U^{(\mS_m)} = r_m^{-1/2} U_m$, where $ r_m = \ol n_m/(N+l)$.
However, in practice, the distributed nodes on the workers are mostly unbalanced
with respect to the whole population.
For instance, smaller samples of the $k$th community may be distributed on the $m$th worker compared to other workers.
As a result, $U^{(\mS_m)}$ will not be just equal to $r_m^{-1/2} U_m$.

This unbalanced effect can be quantified in the theoretical analysis.
Define the {\it unbalanced effect} as $\alpha^{(\mS_m)} = \max_k |\ol n_{mk}/\ol n_m - m_k/N|$.
As a result, $\alpha^{(\mS_m)}$ will be large if the ratio of one community (e.g., the $k$th community)
on the $m$th worker is far away from its population ratio $m_k/N$.
In addition,
let $d_0\le \min_k n_{0k}/l\le \max_k n_{0k}/l \le u_0$
and $d_m\le \min_k n_{mk}/\ol n_m\le \max_k n_{mk}/\ol n_m \le u_m$.
We establish an upper bound for the deviation of $U^{(\mS_m)}$ from $r_m^{-1/2}U_m$.
\bep\label{prop_diff_worker_spectrum}
Let $b_{\min} = \min_{1\le i,j\le K} B_{ij}$.
It holds
\beq\label{equa_worker_spectrum}
\big\|U^{(\mS_{m})}-r_m^{-1/2}U_mQ_m\big\|_F \le  \frac{14\sqrt{2}K^{2}u_m\max\{u_0^{1/2},u_m^{1/2}\}\alpha^{(\mS_m)1/2}}{\sigma_{\min}(B)b_{\min}^3d_0^2d_m^3(d_0+d_m)}+\frac{\alpha^{(\mS_m)}}{d_0}
\eeq
where $Q_m$ is an $K\times K$ orthogonal matrix.
\eep

\noindent
Proof of Proposition \ref{prop_diff_worker_spectrum} is given in Appendix B.3 \ref{Appendix_proof_prop3}.
The upper bound in (\ref{equa_worker_spectrum}) illustrates the relationship between the error bounds and the unbalanced effect.
Particularly, the error bound is tighter when the community members are distributed more evenly on each worker.
In the extreme case, when the unbalanced effect is 0 (i.e., $\alpha ^{(\mS_m)}=0$),
the upper bound in (\ref{equa_worker_spectrum}) will be zero.

%

\csubsection{Convergence of Singular Vectors}

As we have shown previously,
$U^{(\mS_m)}$ has $K$ distinct rows.
As a result, if $\wh U^{(\mS_m)}$ converges to $ U^{(\mS_m)}$ with a high probability,
 we are able to achieve a high clustering accuracy based on spectral clustering using $\wh U^{(\mS_m)}$.
%
In the following theorem we establish the convergence result of $\wh U^{(\mS_m)}$
to $U^{(\mS_m)}$.

\bet\label{thm_U_diff}
{\sc (Singular Vector Convergence)}
Let $\lambda_{1,m}\ge \lambda_{2,m}\ge \cdots \ge\lambda_{K,m}>0$ be the top $K$ singular values of $\mL^{(\mS_m)}$.
Define $\delta_m = \min_i \mD_{ii}^{(\mS_m)}$.
Then for any $\epsilon_m >0$ and $\delta_{m} >
3\log (n_m+2l) + 3\log(4/\epsilon_m)$, with probability at least $1-\epsilon_m$
it holds
\beq
\big\|\wh U^{(\mS_m)} - U^{(\mS_m)}Q^{(\mS_m)}\big\|_F \le \frac{8\sqrt 6}{\lambda_{K,m}}
\sqrt{\frac{K\log(4(n_m+2l)/\epsilon_m)}{\delta_{m}}},\label{upper_U_hat}
\eeq
where $Q^{(\mS_m)}\in \mR^{K\times K}$ is a $K\times K$ orthogonal matrix.
\eet

The proof of Theorem \ref{thm_U_diff} is given in Appendix C.1.
To better understand the estimation error bound given in (\ref{upper_U_hat}),
we make the following comments.
First, the error bound is related to $\lambda_{K,m}$.
According to \cite{rohe2011spectral} and \cite{lei2015consistency},
if $\lambda_{K,m}$ is larger, the eigengap between the eigenvalues
of interest and the rest will be higher.
This enables us to detect communities with higher accuracy level.

Second, the upper bound is lower if the minimum out-degree $\delta_m$ is higher.
One could verify that $\mD_{ii}^{(\mS_m)} = (\Theta_i^{(\mS_m)})^\top B\Theta_0^\top\one_l\ge b_{\min}\sum_k n_{0k} = b_{\min}l$.
Consequently $\delta_m$ grows almost linearly with $l$ if $b_{\min}$ is lower bounded.
If $\delta_m\gg K\log \ol n_m$ and $\lambda_{K,m}$ is lower bounded by a positive constant, then we have
$\|\wh U^{(\mS_m)} - U^{(\mS_m)}Q^{(\mS_m)}\|_F = o_p(1)$.
Lastly, the error bound is higher when the number of communities $K$ and the sub-sample size $\ol n_m$ is larger.
As a result, larger $K$ and $\ol n_m$ will increase the difficulty of the community detection task.

%

\csubsection{Clustering Accuracy Analysis}

In this section, we conduct clustering accuracy analysis for the DCD algorithm.
To this end, we first present a sufficient condition, which guarantees correct clustering for a single node.
Let $\wh C^{(\mS_m)} = (\wh C_{1}^{(\mS_m)},\cdots, \wh C_{K}^{(\mS_m)})^\top\in \mR^{K\times K}$ be the pseudo centers on the worker $m$.
Denote $P_m = (2/D_m)^{1/2} - 2\zeta_m$ with
$D_m=\max_{1\le k\le K}\ol n_{mk}$
and $\zeta_m=\max_{k\in \text{\{1,...,$K$\}}}\|Q^{(\mS_m)\top}U_{i_k}^{(\mS_m)}-\wh C_{k}^{(\mS_m)}\|_2$.
Here $\zeta_m$ characterizes the distance of the pseudo centers to their population values on worker $m$.
We then have the following proposition.
\begin{proposition}\label{prop_clustering_sufficient_cond}
    The node $i$ will be correctly clustered (i.e., $\wh g_i = g_i$) as long as
	\begin{equation}
	\big\|\wh{U}_{i}^{(\mS_m)}-\wh C_{g_i}^{(\mS_m)}\big\|_{2}<\frac{P_m}{2}.\label{condU}
	\end{equation}
\end{proposition}
The proof of Proposition \ref{prop_clustering_sufficient_cond} is given in Appendix B.4.
It indicates that the clustering accuracy is closely related to $P_m$.
If with a high probability that the pseudo nodes are correctly clustered,
then $P_m$ will be higher. As a consequence, it could yield a
higher accuracy of the community detection result.

In the following we analyze the lower bound of $P_m$.
If we could prove that $P_m$ is positive with a high probability, we are then able
to show that the total number of mis-clustered nodes are well controlled.
Specifically,
define the pseudo centers on the master node as $\wh U_{0c} \defeq (\wh U_{0i}:i\in \mC)^\top\in \mR^{K\times K}$.
Ideally, we could directly map $\wh U_{0c}$ to the column space of $\wh U^{(\mS_m)}$
and then complete the community detection on workers.
To this end, a rotation $Q_c$ should be made on the pseudo centers of the master node (i.e., $\wh U_{0c}$).
According to Proposition \ref{prop_diff_worker_spectrum} and Theorem \ref{thm_U_diff},
rotation $Q_c$ takes the form $Q_c = r_m^{-1/2}r_0^{1/2}Q_0^{\top} Q_mQ^{(\mS_m)}$.
As a result, the pseudo centers on the $m$th worker is defined as
$\wh C^{(\mS_m)} = \wh U_{0c} Q_c$.
To establish a lower bound for $P_m$, we first assume the following conditions.
\begin{itemize}
  \item [(C1)] {\sc(Eigenvalue and Eigengap on Master)} Let $\delta_0 = \min_i \mD_{0,ii}$.
  Assume $\delta_0 > 3\log(2l)+3\log(4/\epsilon_{l})$ and $\epsilon_l\rightarrow0$
  as $l\rightarrow\infty$.

  \item [(C2)] {\sc (Pilot Nodes)} Assume $
  K^2\log(l/\epsilon_l)/(b_{\min}\lambda_{K,0}^2)\ll
  l$ with $\epsilon_l\rightarrow0$
  as $l\rightarrow\infty$.
  \item [(C3)] {\sc (Unbalanced Effect)}
  Let $d_0$, $d_m$, $u_0$, $u_m$ be finite constants and assume $\alpha^{(\mS_m)} = o(\sigma_{\min}(B)^2/K^4)$.
\end{itemize}

Condition (C1) is imposed by assuming the same condition as in Theorem \ref{thm_U_diff} for the pilot nodes.
Condition (C2) gives a lower bound on the number of pilot nodes.
Specifically, it should be larger than both the number of communities $K$
and $(\log l)^2/b_{\min}^4$, which is easy to satisfy in practice.

Condition (C3) restricts the unbalanced effect.
First, it states that the relative ratio of communities across all workers are stable by
assuming $d_0$, $d_m$, $u_0$, $u_m$ are constants.
Next, the unbalanced effect $\alpha^{(\mS_m)}$ is assumed to converge to zero faster than $O(1/K^2)$.
As a result, as long as $K$ is well controlled (for instance, in the order of $\log N$) and signal strength in $B$ is strong enough, the conditions (C2) and (C3) could be easily satisfied.
We then have the following Proposition.

\begin{proposition}\label{prop_P_mbiggerthan0}
Assume Conditions (C1)--(C3).
Then with probability $1 - \epsilon_l$, we have $P_m\ge c_1/\sqrt {\ol n_m}$  as $\min\{l, n_m\}\rightarrow\infty$
with rotation $Q_c$,
where $c_1$ is a positive constant.
\end{proposition}

The proof of Proposition \ref{prop_P_mbiggerthan0} is given in Appendix B.5.
In practice, to save us the effort of estimating the rotation matrix $Q_c$,
we directly broadcast the pseudo center indexes $\mC$ to workers and let $\wh C^{(\mS_m)} = (\wh U_i^{(\mS_m)}: i\in \mC)^\top$.
As a result, $\wh C^{(\mS_m)}$ is naturally embedded in the column space of $\wh U^{(\mS_m)}$ and no further rotation is required.
Given the results presented in Theorem \ref{thm_U_diff} and Proposition \ref{prop_P_mbiggerthan0}, we are then able to obtain the
mis-clustering rates for each worker as follows.

\bet\label{thm_misclustered_nodes}
{\sc (Bound of mis-clustering Rates)} Assume conditions in Theorem \ref{thm_U_diff} and Proposition \ref{prop_P_mbiggerthan0}.
Denote $ \cR^{(\mS_m)}$ as the ratio of misclustered nodes on worker $m$, then we have
\beq
    \cR^{(\mS_m)} =o\left(\frac{K^2\log(l/\epsilon_l)}{b_{\min}l\lambda_{K,0}^2}+\frac{K\log(4(n_m+2l)/\epsilon_m)}{\lambda_{K,m}\delta_m}+\frac{K^4\alpha^{(\mS_m)}}{\sigma_{\min}(B)^2b_{\min}^6}\right), \label{misclustered_number_discuss}
\eeq
with probability at least $1-\epsilon_l-\epsilon_m$.
\eet

 The proof of Theorem \ref{thm_misclustered_nodes} is given in Appendix C.2.
	Theorem \ref{thm_misclustered_nodes} establishes an upper bound for the mis-clustering rate on the worker $m$.
With respect to the result, we have the following remark.

\begin{remark}
One could observe that there are three terms included in the mis-clustering rate.
The first and second terms are related to convergence of spectrum on master and workers.
Specifically, the first term is related to the convergence of eigenvectors on the master.
The second term is determined by convergence of singular vectors on the $m$th worker.
As we comment before, with large sample size and strong signal strength,
the mis-clustering rate could be well controlled.
Next,
the third term is mainly related to the unbalanced effect $\alpha^{(\mS_m)}$ among the workers,
which is lower if the distribution of the communities is more balanced on different workers.
\end{remark}

{
Compared with using the full adjacency matrix of $\mS_m$,
the error bound in (\ref{misclustered_number_discuss}) is higher.
That is straightforward to understand since in our case we use a sub-adjacency matrix instead of the full one.
 According to \cite{rohe2012co}, when the full adjacency matrix is used, the mis-clustering rate is bounded by $\cR^{(\mS_m)}_{all}= O(K\log(\ol n_m/\epsilon_m)/(\ol n_m\lambda_{K,m}^2))$ with high probability.
In our case, we have $\cR^{(\mS_m)} = O(\cR^{(\mS_m)}_{all}\ol n_m/l)$.
Hence if it holds $n_m\ll l$ (i.e., $\ol n_m\approx l$), then the mis-clustering rate is asymptotically the same as using the full adjacency matrix.
Furthermore, we can obtain a mis-clustering error bound for all network nodes as in the following corollary.
\begin{corollary}\label{coro1}
  Assume the same conditions as in Theorem \ref{thm_misclustered_nodes}.
  In addition, assume $n_1 = n_2=\cdots = n_M \defeq n$ and $\alpha^{(\mS_m)} = 0$ for $1\le m\le M$.
  Denote $\cR_{all}$ as number of all mis-clustered nodes across all workers.
  Then with probability $1-(M+1)/l$ we have
  \begin{align}
  \cR_{all} = O\Big(\frac{K(\log n + \log l)}{l\lambda_{K}^2}\Big),\label{e_all_bound}
  \end{align}
  where $\lambda_K = \min_m \lambda_{K,m}$.
\end{corollary}
The Corollary \ref{coro1} could be immediately obtained from Theorem \ref{thm_misclustered_nodes}
by setting $\epsilon_l =\epsilon_m = 1/l$ for $1\le m\le M$.
As indicated by (\ref{e_all_bound}), the mis-clustering rate is smaller when the number of pilot nodes $l$ is larger.
Particularly, if $l = rN$ with $r\in (0,1)$ being a finite positive constant, then the mis-clustering rate is almost the same as we use the whole adjacency matrix $A$.
While in the same time, the computational time is roughly $r^2$ smaller than using the whole adjacency matrix.
As a consequence, the computational advantage is obvious.

\begin{remark}
  We compare our theoretical findings with \cite{yang2015divide} and \cite{mukherjee2017two}.
  The mis-clustering rate of  \cite{mukherjee2017two} is comparable to our method.
  That is, to achieve the same global mis-clustering rate, the required subgraph size of \cite{mukherjee2017two} is in the same order as we require for our pilot network.
  \cite{yang2015divide}   could allow for smaller subgraph sizes to obtain the same global mis-clustering rate.
  However, we find that they require a denser network setting, which  requires the network density to be $O({\rm{poly}(\log N)}/{N^{\alpha}})$ for some $0<\alpha<1$,
  where ${\rm{poly}(\log N)}$ denotes a polynomial order of $\log N$.
  As an alternative, we could allow for a sparser network setting by allowing $\alpha = 1$.
  We compare the finite sample performances with \cite{yang2015divide}  under various network settings in the numerical study.
\end{remark}

 \csection{SIMULATION STUDIES}\label{section_simulations}

 \csubsection{Simulation Models and Performance Measurements}

 In order to demonstrate the performance of our DCD algorithm, we conduct experiments using synthetic datasets under three scenarios.
 The main differences lie in the generating mechanism of the networks.
 For simplicity, we consider a stochastic block model with $K$ blocks and each block contains $s$ nodes.
 As a result, $Ks = N$.
 The connectivity matrix $B$ is set as
 \beq
 B = \nu\big\{\lambda I_K + (1-\lambda)\one_K\one_K^\top\big\},\label{simu_B}
 \eeq
 where $\nu\in [0,1]$ and $\lambda \in [0,1]$.
 By (\ref{simu_B}), the connection intensity is then parameterized by $\nu$ and
 the connection divergence is characterized by $\lambda$.

 The random experiments are repeated for $R = 500$ times for a reliable evaluation.
 To gauge the finite sample performance, we consider two accuracy measures.
 The first is the mis-clustering rate, i.e., $\cR_{all} = \sum_{i = 1}^{N}(\wh g_i\ne g_i)/N$.
 The second is the estimation accuracy of the singular vectors, i.e., $\wh U^{(\mS_m)}$ for each worker,
 which is captured by the log-estimation error (LEE).
  Define $\mbox{LEE}_m = \log\|\wh U^{(\mS_m)} - U^{(\mS_m)}Q^{(\mS_m)}\|_F$
   for the $m$th worker, where the rotation matrix $Q^{(\mS_m)}$ is calculated according to \cite{rohe2011spectral}.
 Subsequently $\mbox{LEE} = M^{-1}\sum_m \mbox{LEE}_m$ is calculated to quantify the average estimation errors over all workers.

 \csubsection{Simulation Results}

%
%

{\sc Scenario 1 (Pilot Nodes)}
First, we investigate the role of pilot nodes on the numerical performances.
Particularly, we let $l=rN$ with $N=10000$ and $r$ varying from $0.01$ to $0.2$.
The performances are evaluated for $K=3,4,5,6$ and
the connection intensity and divergence are fixed as $\nu = 0.2$, $\lambda = 0.5$.
In addition, the number of workers is given as $M = 5$.
We calculated the mis-clustering rate $\cR_{all}$ in the left panel of Figure \ref{fig_scenario_1}.
As shown in Figure \ref{fig_scenario_1}, the mis-clustering rate converges to zero as $l$ grows, which corroborates with our theoretical findings in Corollary \ref{coro1}.

\begin{figure}[htb]
	\centering
		\includegraphics[width=\textwidth]{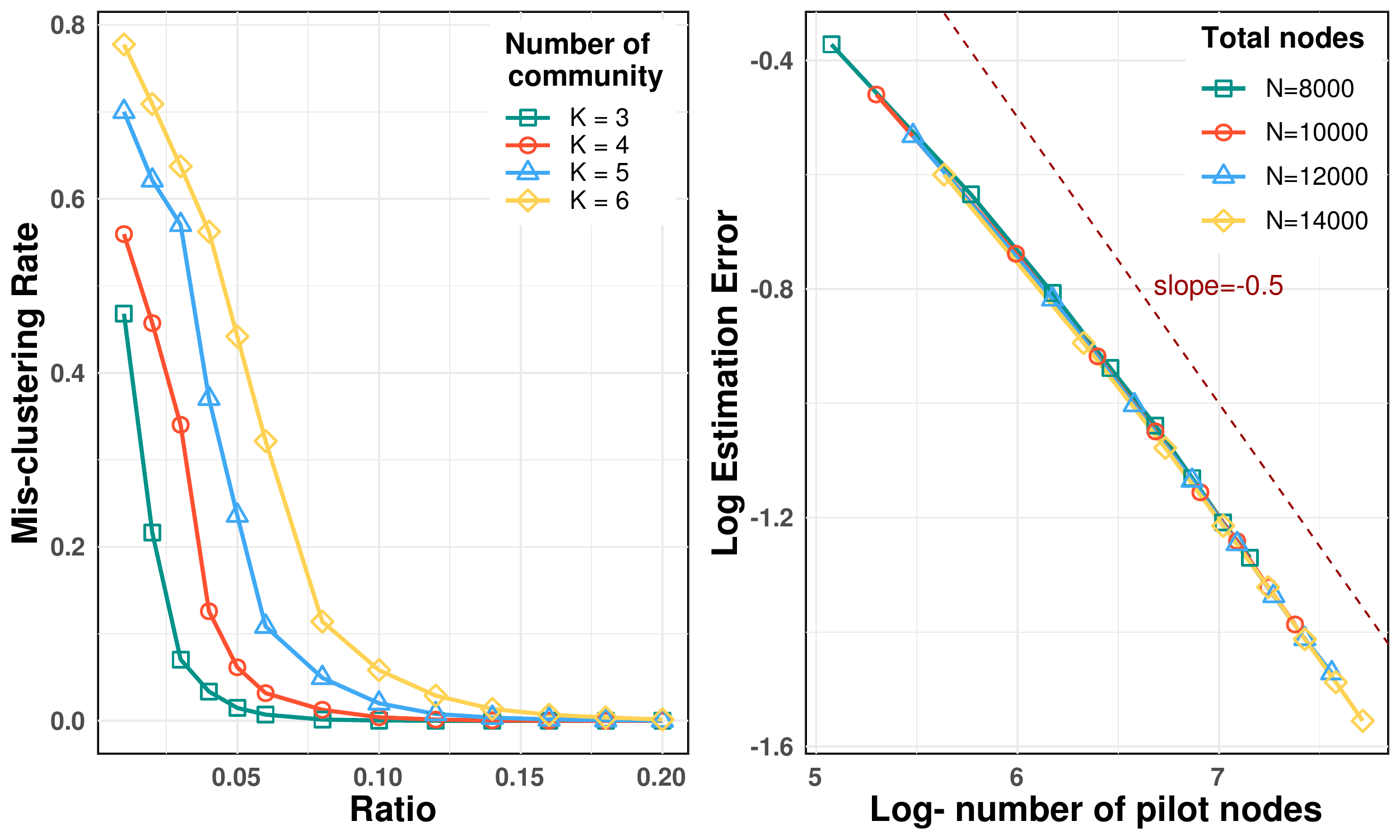}
	\caption{Left panel: the mis-clustering rates versus pilot nodes ratio (i.e., $l/N$) under different community sizes $K = 3,4,5,6$;
Right panel: LEE versus the log-number of pilot nodes under sample sizes $N = 8000,10000, 12000, 14000$.
}
	\label{fig_scenario_1}
\end{figure}

Moreover, we evaluate the estimation accuracy of the estimated eigenvectors by LEE.
As we can see from the right panel of Figure  \ref{fig_scenario_1}, for a fixed $N$, as $\log(l)$ grows, the estimation error of eigenvectors decreases with the slope of LEE roughly parallel with $-1/2$.
This corroborates with the theoretical results in Theorem \ref{thm_U_diff}.

\begin{figure}[htb]
	\centering
	\includegraphics[width=\textwidth]{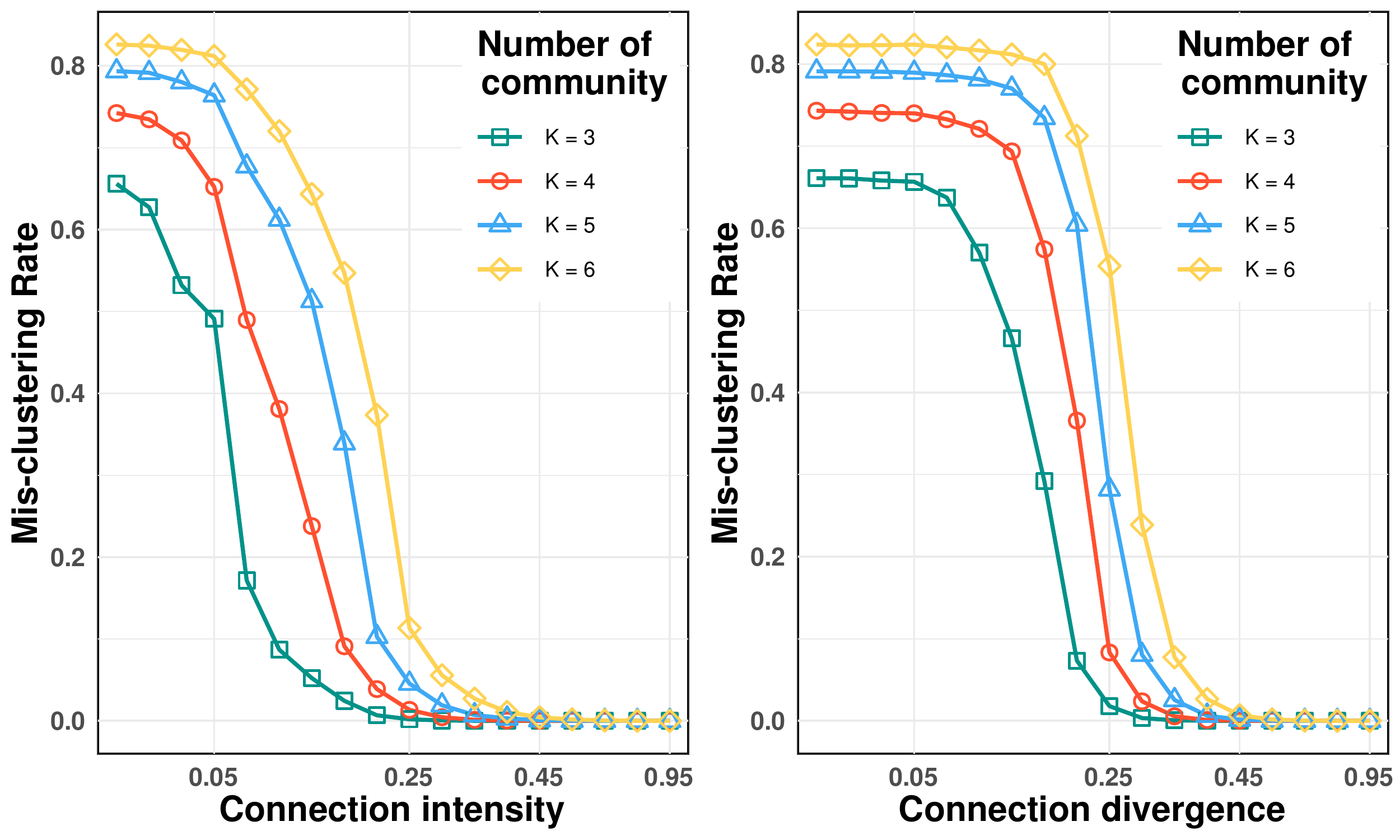}
	\caption{The influence of $\nu$ (connection intensity) and $\lambda$ (connection divergence)
on the mis-clustering rates.
As shown in the figure, stronger connection and larger divergence can lead to more accurate community detection results. }
	\label{fig3_1}
\end{figure}

 {\sc Scenario 2 (Signal Strength)}
 In this scenario, we observe how the mis-clustering rates change with respect to the signal strengths.
 Accordingly we fix $l = 300$, $N = 20000$, and vary the number of communities as $K = 3,4,5,6$.
 For $\nu=0.2$, we first change the connection divergence from $\lambda = 0.05$ to $\lambda = 0.95$.
 As $\lambda$ increases, the connection intensity within the same community will be higher than nodes from different communities.
 In the meanwhile, the eigengap is larger and the signal strength is higher.
 Next, we conduct the experiment {by varying connection intensity $\nu $  and fix $\lambda = 0.5$.
 Theoretically, as $\nu$ increases, $b_{\min}$ will increase accordingly, which results in a higher signal strength.
 According to Theorem \ref{thm_U_diff} and Corollary \ref{coro1}, the mis-clustering rates will drop as the signal strength is higher.
 This phenomenon can be confirmed from the right panel of Figure \ref{fig3_1}.

{\sc Scenario 3 (Unbalanced Effect)}
 In this setting, we verify the unbalanced effect on the finite sample performances.
 First, we fix $l = 500$, $N = 5000$, $\nu = 0.2$, $\lambda = 0.5$, and $M=3$.
 Denote $\pi_{mk}$ as the ratio of nodes in the $k$th community on the $m$th worker.
 We set $\pi_{mk}$ as follows,
  \beq
 \pi_{mk} = \frac{1}{K} + \Big(k - \frac{K+1}{2}\Big)\mbox{sign}\Big(m - \frac{M+1}{2}\Big)\frac{\alpha}{K(K-1)}.\nonumber
 \eeq
 If $\pi_{m1} = \pi_{m2}=\cdots = \pi_{mK} = 1/K$, then there is no unbalanced effect.
 As $\alpha$ increases, the unbalanced effect is larger.
 The mis-clustering rate is visualized in Figure \ref{fig_Scenario4}.
 As $\alpha$ is increased from $0$ to $0.95$, we could observe that the
 mis-clustering rates increase accordingly, which verifies the result of Theorem \ref{thm_misclustered_nodes}.

\begin{figure}[htb]
	\centering
	\includegraphics[width=0.7\textwidth]{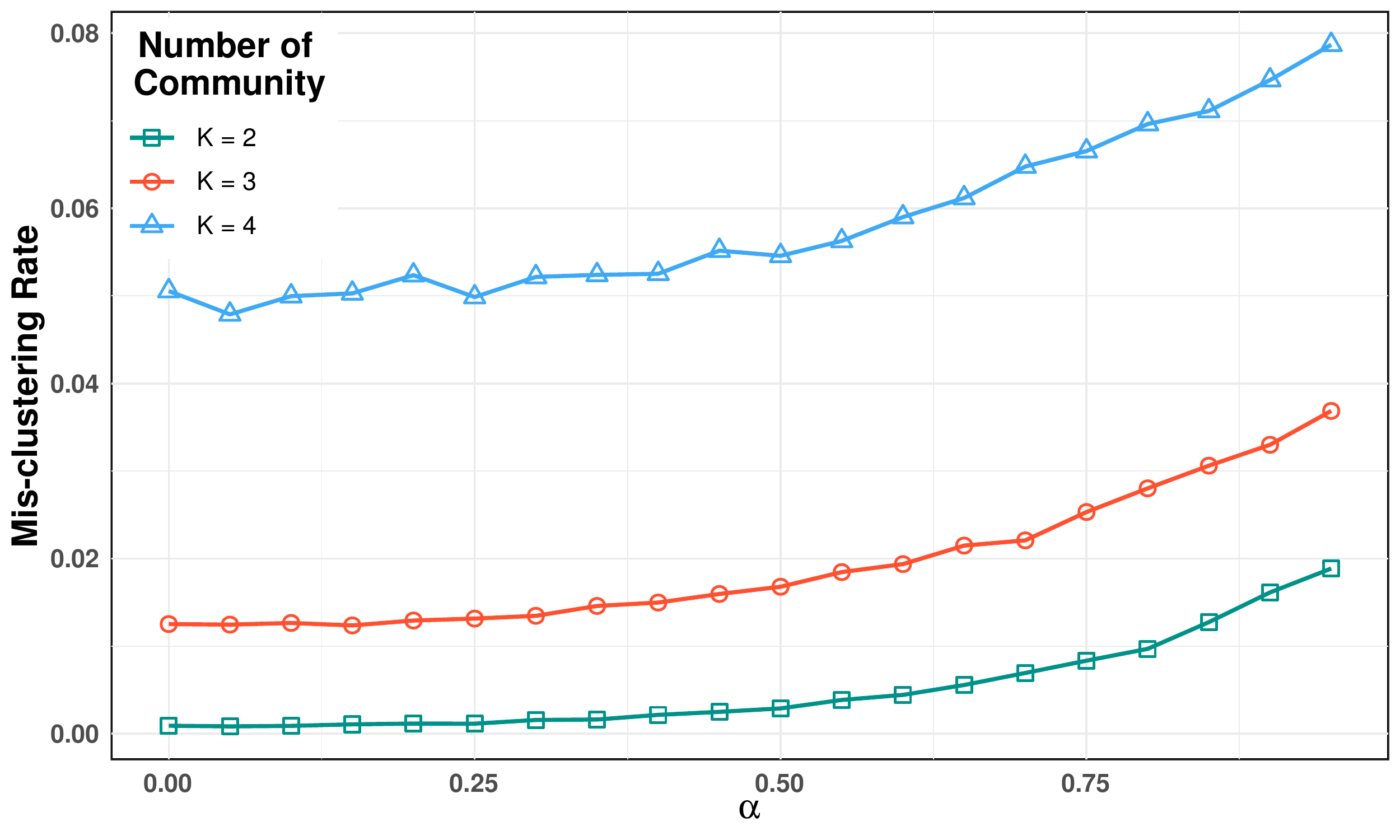}
	\caption{The mis-clustering rates versus the unbalanced effect $\alpha$ for different number of communities $K=2, 3, 4$. As the unbalanced effect increases, the mis-clustering rates also increase, which results in inferior performance of the distributed algorithm.
}
	\label{fig_Scenario4}
\end{figure}

\csubsection{Numerical Comparisons}

Lastly, we illustrate the powerfulness of our method by making numerical comparisons with existing approaches in terms of clustering accuracy and computational efficiency.
Specifically, the conventional spectral clustering (SC) method and the distributed graph clustering (DGC) method proposed by \cite{yang2015divide} are included. The details are given as follows.

\scsubsection{Comparison with SC Method}

First, we compare the performances of the proposed method with the SC method using the whole network data.
For a network with size $N$, we conduct spectral clustering in Algorithm \ref{sc_all}
and record the clustering accuracy and computational time.
For comparison, we conduct distributed spectral clustering using $M$ workers by Algorithm \ref{distribute_SBM}.

\begin{figure}[htb]
	\centering
	\includegraphics[width=\textwidth]{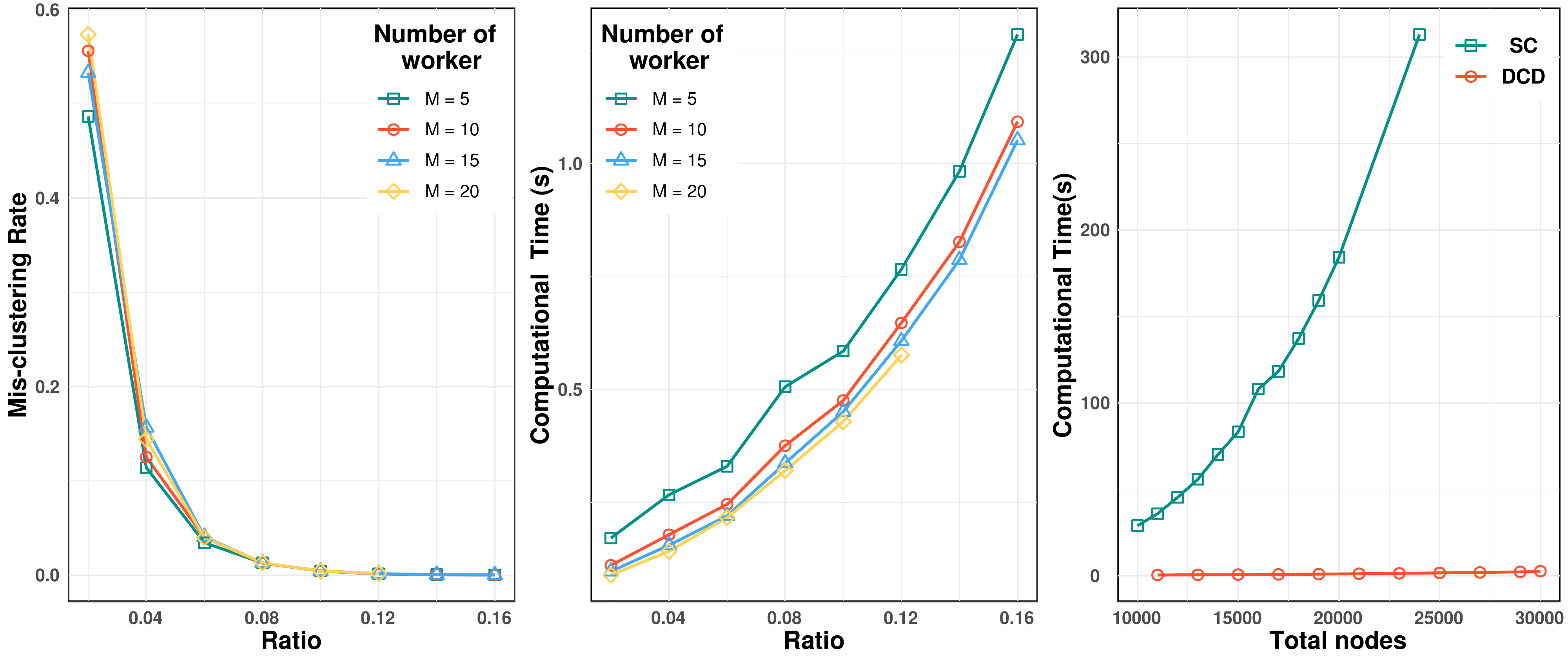}
	\caption{The mis-clustering rates (left panel) and computational time (middle panel) with respect to
varying pilot nodes ratios for different number of workers.
The computational times of SC Algorithm \ref{sc_all} and DCD Algorithm \ref{distribute_SBM} are further compared as $N$ grows (right
panel).}
	\label{fig_comparison1}
\end{figure}

For $N=10000$ and $K=3$, the average computational time of spectral clustering Algorithm \ref{sc_all} (using the whole network adjacency matrix) is 48.84s and the mis-clustering rate is zero.
Next, we set $l=rN$ with $r=0.02,0.04,...,0.18$ for Algorithm \ref{distribute_SBM}. Both mis-clustering rates and computational time are compared,
which is shown in Figure \ref{fig_comparison1}.
As we could observe, after $l \ge 0.10\times10000 = 1000$, our algorithm could obtain the same mis-clustering rate but with much lower computational cost (less than 1 second).
In addition, with more workers, the computational cost will be further reduced.
Lastly, we compare the computational times as $N$ grows.
For each $N$, $l$ is set when the mis-clustering rate is the same as using the whole adjacency matrix.
As we can observe from the right panel of Figure \ref{fig_comparison1},
as the network size grows, the computational time of Algorithm \ref{sc_all} increases drastically compared to Algorithm \ref{distribute_SBM},
which illustrates the computational advantage of the proposed approach.

\scsubsection{Comparison with DGC Method}

Next, we compare with the DGC method of \cite{yang2015divide}.
Specifically, we consider both dense and sparse network settings by setting $\nu  = 0.6,\ \lambda=0.5$ and $\nu=0.2,\ \lambda=0.5$ in (\ref{simu_B}) respectively.
For both settings, we fix the number of workers $M = 10$ and let $N = 1000, 2000, \cdots, 10^4$.
For the DCD method, we set the pilot network size as $l = 0.1N$.
For the DGC method, we set the subgraph size as $N/M$
and use grid search method to tune all the hyper-parameters.
The simulation results are given in Figure \ref{fig_comparison2}.

\begin{figure}[htb]
	\centering
	\includegraphics[width=\textwidth]{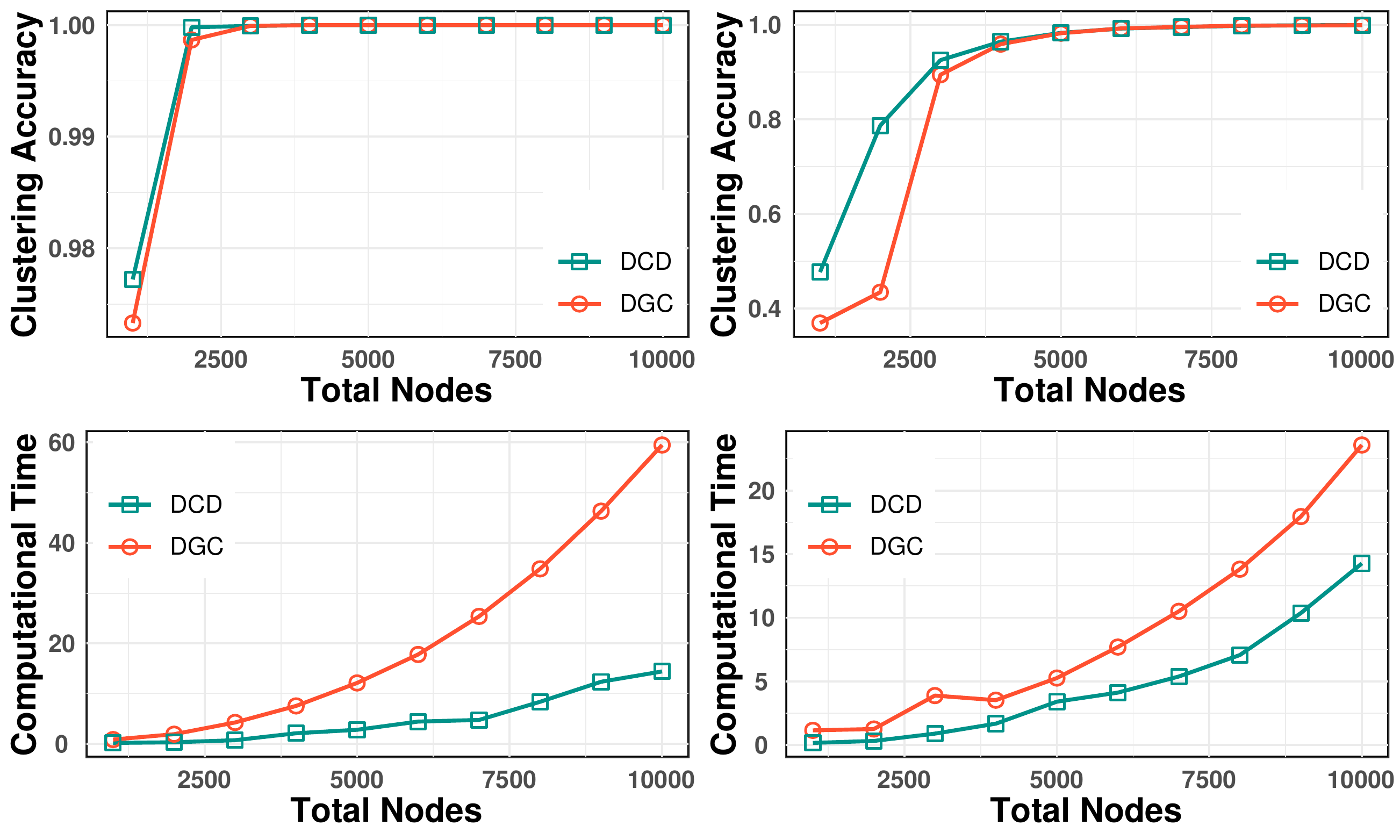}
	\caption{Left panel: the clustering accuracy and computational time on a dense network (i.e., $\nu=0.6$) under different total nodes $N=1000, 2000, \ldots, 10^4$; Right Panel: the clustering accuracy and computational time on a sparse network (i.e., $\nu=0.2$) under different total nodes $N=1000, 2000, \ldots, 10^4$.}
	\label{fig_comparison2}
\end{figure}

For both dense and sparse network settings, we could observe that the DCD method is able to achieve a higher clustering accuracy with relatively lower computational cost.
For the dense network setting, both methods obtain 100\% accuracy when $N \ge  3000$.
However, we observe that as $N$ increases, the computational cost of DGC method grows much faster than the DCD method.
For the sparse network setting, the DCD method has clearly higher clustering accuracy rate especially when $N$ is small.
This corroborates the stringent theoretical requirement of the network density given in \cite{yang2015divide}.

\csection{EMPIRICAL STUDY}

We evaluate the empirical performance of the proposed method using two network datasets.
The estimation accuracy and computational time are evaluated using both distributed community detection algorithm and spectral clustering
method.
Particularly, the distributed community detection algorithm is implemented using our newly developed package \textsf{DCD}
on the Spark system.
The system consists {36} virtual cores and {128} GB of RAM.
We set the number of workers as {M=2}.
Descriptions of the two network datasets and corresponding experimental results are presented as follows.

%

\csubsection{Pubmed Dataset: a Citation Network}

The Pubmed dataset consists of 19,717 scientific publications from PubMed database \citep{kipf2016semi}.
Each publication is identified as one of the three classes, i.e., Diabetes Mellitus Experimental,
Diabetes Mellitus Type 1, Diabetes Mellitus Type 2.
The sizes of the three classes are 4,103, 7,875, and 7,739 respectively.
In this case the community sizes are relatively unbalanced since both the second and third classes have roughly twice members than the first class.
The network link is defined using the citation relationships among the publications.
Specifically, if the $i$th publication cites the $j$th one (or otherwise), then $A_{ij} = 1$, otherwise $A_{ij} = 0$.
The resulting network density is $0.028\%$.

For the Pubmed datasets, we could calculate the mis-clustering rates by using pre-specified class labels  as  ground truth.
The mis-clustering rates of using SC with the whole network are {$33.03\%$}.
For comparison, the DCD algorithm is evaluated
by varying $r=l/N $ from 0.02 to 0.30.
One could observe in Figure \ref{fig_pubmed} that, the mis-clustering rates of the DCD algorithm is comparable to the
SC algorithm when $r = 0.22$, {while the computational time is much lower}.

\begin{figure}[htb]
	\centering
	\includegraphics[width=1\textwidth]{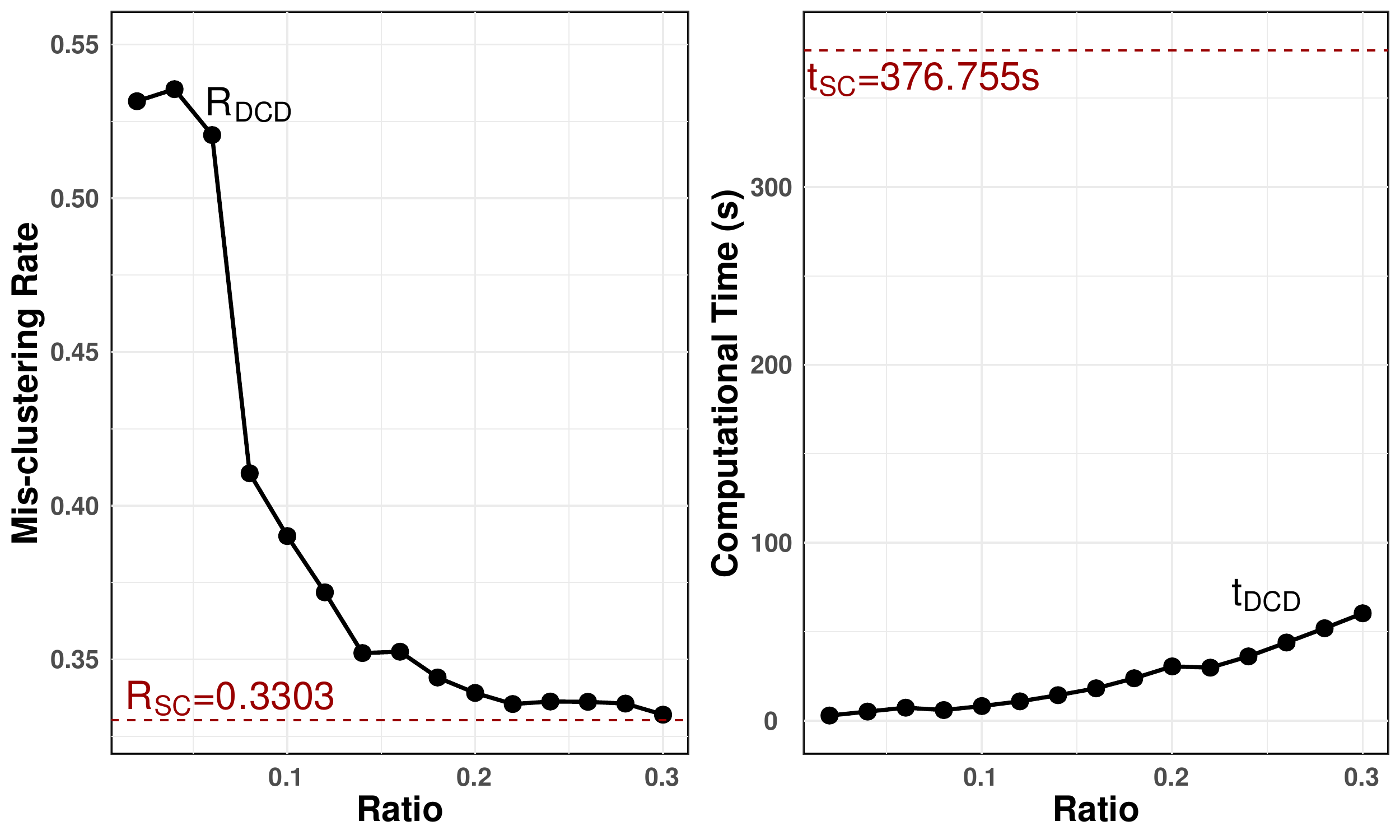}
	\caption{The Comparison between SC algorithm and DCD algorithm on Pubmed dataset both in mis-clustering rate and computational time.
The mis-clustering rate and computational time of the DCD (SC) algorithm are denoted by
R$_{\rm DCD}$ (R$_{\rm SC}$) and t$_{\rm DCD}$ (t$_{\rm SC}$) respectively.}
	\label{fig_pubmed}
\end{figure}

\csubsection{Pokec Dataset: an Online Social Network}

In this study, we consider a large scale social network, Pokec \citep{takac2012data}.
The Pokec is the most popular online social network in Slovak.
The dataset was collected during May 25--27 in the year of 2012,
which contains 50,000 active users in the network.
If the $i$th user is a friend of the $j$th user, then there is a connection between the two users, i.e.,
$A_{ij} = 1$.
The resulting network density is {$0.0985\%$}.


%

For the Pokec dataset, the network size is too huge for SC algorithm to output the result.
As an alternative, we perform our DCD algorithm to conduct community detection.
Since the memberships are not available, we produce
another clustering criterion instead.
Specifically, define the relative density as RED$=\mbox{Den}_{between}/\mbox{Den}_{within}$,
where $\mbox{Den}_{between} = \sum_{i,j}a_{i,j}I(\wh g_i\ne \wh g_j)/\sum_{i,j} I(\wh g_i\ne \wh g_j)$ is the between-community density,
and $\mbox{Den}_{within} = \sum_{i,j}a_{i,j}I(\wh g_i= \wh g_j)/\sum_{i,j} I(\wh g_i= \wh g_j)$ is the within-community density.
The RED is visualized in Figure \ref{fig_pokec}.
As one can observe, after $l/N \ge 0.28$, the RED is stable with the corresponding computational time as 642.524s.
This further illustrates the computational advantage of the proposed DCD algorithm.

\begin{figure}[H]
	\centering
	\includegraphics[width=0.8\textwidth]{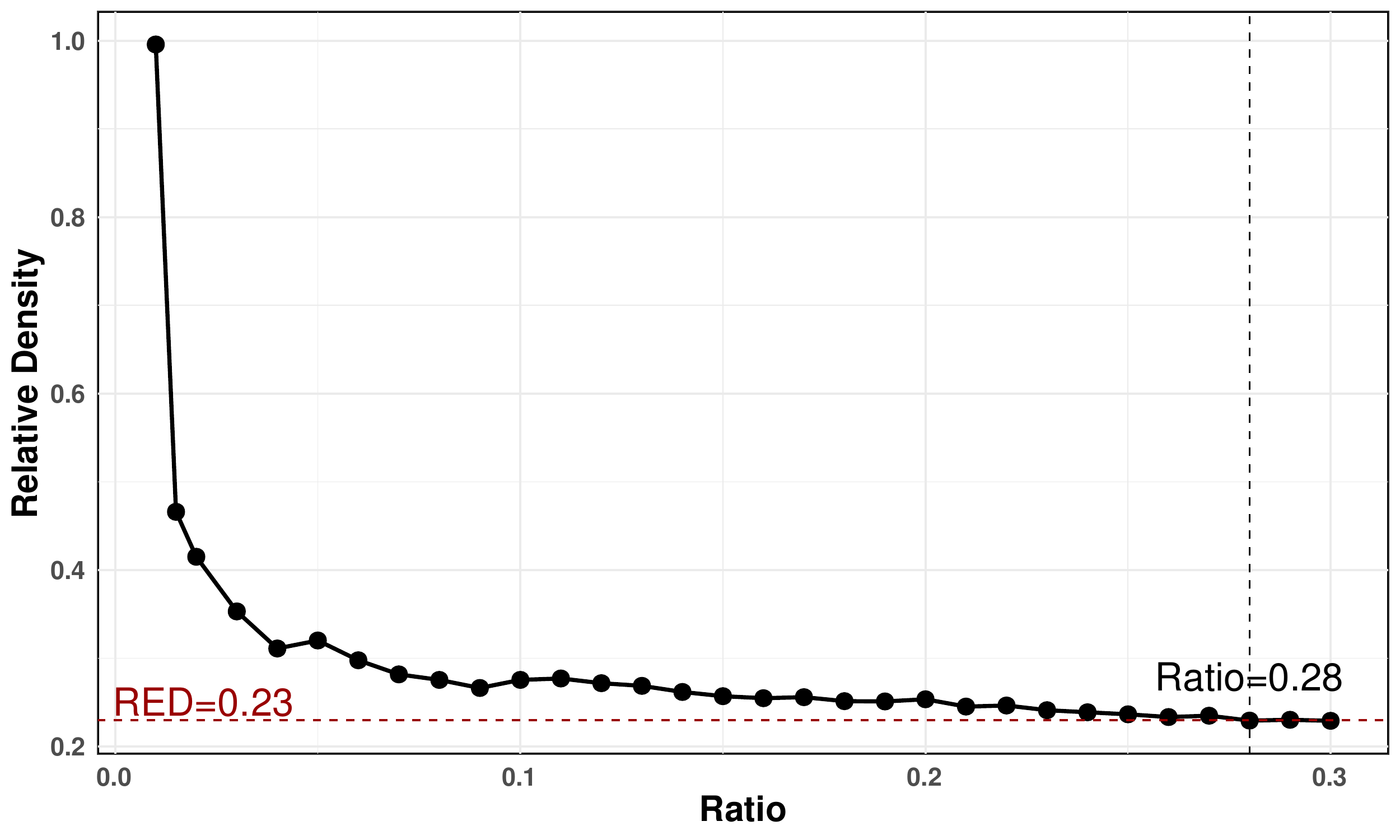}
	\caption{The relative density decreases rapidly as ratio increases, after $r = l/N \ge 0.28$, the RED is stable around 0.23.}
	\label{fig_pokec}
\end{figure}

\csection{CONCLUDING REMARKS}

In this work, we propose a distributed community detection (DCD) algorithm to tackle community detection task
in large scale networks.
We distribute $l$ pilot nodes on the master and a non-square adjacency matrix on workers.
The proposed DCD algorithm has three merits.
First, the communication cost is low.
Second, no further iteration algorithm is used on workers therefore the algorithm is stable.
Third, both the computational complexity and the storage requirements are much lower compared to using the whole adjacency matrix.
The DCD algorithm is shown to have clear computational advantage and competitive statistical performance by using a variety synthetic and empirical datasets.

To conclude the article, we provide several topics for future studies.
First, better mechanisms can be designed to select pilot nodes on the master server.
This enables us to obtain more accurate estimation of the pseudo centers and yields better clustering results.
Next, it is interesting to extend the proposed method to directed network by considering sending and receiving clusters respectively \citep{rohe2012co}.
The theoretical property and computational complexity could be discussed accordingly.
Third, in the community detection task, we only employ the network structure information and ignore other potential useful nodal covariates.
As a result, it is important to extend the DCD algorithm to further incorporate
various exogenous information.

\bibliographystyle{asa}
\bibliography{reference}

\newpage
\scsection{APPENDIX A}
\renewcommand{\theequation}{A.\arabic{equation}}
\setcounter{equation}{0}

\begin{appendices}

\scsubsection{Appendix A.1: Notations, Useful Lemmas and Propositions}

We first define several notations which will be used in the rest of the proof.
Let
\beq
\wt L^{(\mS_m)}= \left(
\begin{array}{cc}
\zero_{\ol n_m\times \ol n_m} &L^{(\mS_m)}\\
L^{(\mS_m)\top}& \zero_{l\times l}\end{array}
\right),~~
\wt \mL^{(\mS_m)}= \left(
\begin{array}{cc}
\zero_{\ol n_m\times \ol n_m} &\mL^{(\mS_m)}\\
\mL^{(\mS_m)\top}& \zero_{l\times l}\end{array}
\right).\nonumber
\eeq
Similarly, define $\wt A^{(\mS_m)}, \wt \bA^{(\mS_m)}\in \mR^{(n_m+2l)\times (n_m+2l)}$
in the same way.
In addition, define $\wt D^{(\mS_m)} = \diag\{D^{(\mS_m)}, F^{(\mS_m)}\}$
and $\wt \mD^{(\mS_m)} = \diag\{\mD^{(\mS_m)}, \mF^{(\mS_m)}\}$.

\bep\label{prop_L_bound}
Assume the same conditions with Theorem \ref{thm_U_diff}.
Then we have
\beq
\big\|\wt L^{(\mS_m)} - \wt\mL^{(\mS_m)}\big\|_{\max}\le 4\sqrt3
\sqrt{\frac{\log(4(n_m+2l)/\epsilon_m)}{
\delta_{m}}}.\label{L_diff}
\eeq
with probability at least $1-\epsilon_m$.
\eep

\begin{proof}

Note that $\wt D^{(\mS_m)}$ and $\wt L^{(\mS_m)}$ are dependent with each other.
Therefore, we add an intermediate step by using the matrix
$\wt C^{(\mS_m)}\defeq (\wt\mD^{(\mS_m)})^{-1/2}\wt A^{(\mS_m)}
(\wt\mD^{(\mS_m)})^{-1/2}$.
Hence we have
\beq
\big\|\wt L^{(\mS_m)} - \wt\mL^{(\mS_m)}\big\|_{\max}\le
\big\|\wt L^{(\mS_m)} - \wt C^{(\mS_m)}\big\|_{\max} +
\big\|\wt \mL^{(\mS_m)} - \wt C^{(\mS_m)}\big\|_{\max}.
\eeq
Define $\nu = \sqrt{3 \log(4(n_m+2l)/\epsilon_m)/\delta_{m}}$.
Then we have $\nu \le 1$ for sufficiently large $l$.
By (\ref{L_C_diff}) and (\ref{L_C_diff2})
of Proposition \ref{lem_L_C_diff}, we have
$\|\wt L^{(\mS_m)} - \wt\mL^{(\mS_m)}\|_{\max}\le
\nu^2 + 3\nu \le 4\nu$.
This yields (\ref{L_diff}).


\end{proof}

\bep\label{lem_L_C_diff}
Assume the same conditions in Theorem \ref{thm_U_diff}.
Let $\wt C^{(\mS_m)}\defeq (\wt\mD^{(\mS_m)})^{-1/2}\wt A^{(\mS_m)}\\
(\wt\mD^{(\mS_m)})^{-1/2}$.
Then we have with probability at least $1-\epsilon_m/2$
\begin{align}
&\big\|\wt L^{(\mS_m)} - \wt C^{(\mS_m)}\big\|_{\max} \le \nu^2+2\nu,\label{L_C_diff}\\
&\big\|\wt \mL^{(\mS_m)} -\wt C^{(\mS_m)}\big\|_{\max} \le \nu,\label{L_C_diff2}
\end{align}
where
\beq
\nu = \sqrt{3 \log\{4(n_m+2l)/\epsilon_m\}/
	\delta_{m}}.                \label{nu_def}
\eeq
\eep

\begin{proof}
	Note that $\wt\mD^{(\mS_m)}_{ii}\ge \delta_m$ for $i=1,...,\ol n_m+l$. We prove (\ref{L_C_diff}) and (\ref{L_C_diff2}) respectively as follows.
	
	\noindent
	{\sc 1. Proof of (\ref{L_C_diff}).}\\
		It can be derived that
		\begin{align*}
		&\big\|\wt{L}^{(\mS_m)}-\wt{C}^{(\mS_m)}\big\|_{\max} =
		\big\|\wt{L}^{(\mS_m)}-(\wt{\mathcal{D}}^{(\mS_m)})^{-1 / 2} (\wt{D}^{(\mS_m)})^{1 / 2} \wt{L}^{(\mS_m)} (\wt{D}^{(\mS_m)})^{1 / 2} (\wt{\mathcal{D}}^{(\mS_m)})^{-1 / 2}\big\|_{\max}\\
		& \le \Big\|\big\{I-(\wt{\mathcal{D}}^{(\mS_m)})^{-1 / 2} (\wt{D}^{(\mS_m)})^{1 / 2}\big\} \wt{L}^{(\mS_m)} (\wt{D}^{(\mS_m)})^{1 / 2} (\wt{\mD}^{(\mS_m)})^{-1 / 2}\Big\|_{\max}\\
		&+ \Big\|\wt{L}^{(\mS_m)}\big\{I-(\wt{D}^{(\mS_m)})^{1 / 2} (\wt{\mathcal{D}}^{(\mS_m)})^{-1 / 2}\big\}\Big\|_{\max}\defeq \Delta_1 + \Delta_2.
		\end{align*}
		We then deal with $\Delta_1$ and $\Delta_2$ respectively.
		Note that we have $\|\wt{L}^{(\mS_m)}\|_{\max} \le 1$
		and $\wt D^{(\mS_m)}$, $\wt\mD^{(\mS_m)}$ are diagonal matrices.
		Hence we have
		$\Delta_2 \le \big\|I-(\wt{\mathcal{D}}^{(\mS_m)})^{-1 / 2} (\wt{D}^{(\mS_m)})^{1 / 2}\big\|_{\max}$ and
		\begin{align*}
		\Delta_1 &\le  \big\|I-(\wt{\mathcal{D}}^{(\mS_m)})^{-1 / 2} (\wt{D}^{(\mS_m)})^{1 / 2}\big\|_{\max}\big\|
		(\wt{\mathcal{D}}^{(\mS_m)})^{-1 / 2} (\wt{D}^{(\mS_m)})^{1 / 2}\big\|_{\max}\\
		&\le \big\|I-(\wt{\mathcal{D}}^{(\mS_m)})^{-1 / 2} (\wt{D}^{(\mS_m)})^{1 / 2}\big\|_{\max}^2 + \big\|I-(\wt{\mathcal{D}}^{(\mS_m)})^{-1 / 2} (\wt{D}^{(\mS_m)})^{1 / 2}\big\|_{\max}.
		\end{align*}
		Then it suffices to bound
		$\big\|I-(\wt{\mathcal{D}}^{(\mS_m)})^{-1 / 2} (\wt{D}^{(\mS_m)})^{1 / 2}\big\|_{\max}$.
		By \cite{chung2006complex},
		it holds that
		\beq
		\mathbb{P}
		\left(\big| \wt{D}^{(\mS_m)}_{i i}-\wt{\mathcal{D}}^{(\mS_m)}_{i i} \big| \geq \lambda\right) \leq \exp \Big(-\frac{\lambda^{2}}{2\wt{\mathcal{D}}^{(\mS_m)}_{i i}}\Big)+
		\exp \Big(-\frac{\lambda^{2}}{2\wt{\mathcal{D}}^{(\mS_m)}_{i i}+2\lambda/3}\Big).\label{concen_ine}
		\eeq
		Note that,
		\begin{align*}
		\Big\|(\wt{\mathcal{D}}^{(\mS_m)})^{-1 / 2} (\tilde{D}^{(\mS_m)})^{1 / 2}-I\Big\|_{\max}&=\max_{i} \big| (\wt D_{ii}^{(\mS_m)})^{1/2}(\wt\mD_{ii}^{(\mS_m)})^{-1/2} -1\big|\\
		&\leq
		\max _{i}\big|\wt{D}^{(\mS_m)}_{i i}/\wt\mD_{ii}^{(\mS_m)}-1\big|
		\end{align*}
		
		This implies for any fixed $\nu$,
		\begin{align*}
		&\mathbb{P}\left(\big\|(\wt{\mathcal{D}}^{(\mS_m)})^{-1 / 2} (\wt{D}^{(\mS_m)})^{1 / 2}-I\big\|_{\max} \geq \nu\right)  \leq \mathbb{P}\left(\max_{i}\big| \wt{D}^{(\mS_m)}_{i i}/\wt\mD_{ii}^{(\mS_m)}-1 | \geq \nu\right) \\
		& \le \sum_i \mP\Big\{\Big|
		\wt{D}^{(\mS_m)}_{i i} - \wt{\mD}^{(\mS_m)}_{i i}
		\Big|\ge \nu\wt{\mD}^{(\mS_m)}_{i i}
		\Big\}
		\end{align*}
		By using (\ref{concen_ine}) we further have
		\begin{align*}
		&\mP\Big\{\Big|
		\wt{D}^{(\mS_m)}_{i i} - \wt{\mD}^{(\mS_m)}_{i i}
		\Big|\ge \nu\wt{\mD}^{(\mS_m)}_{i i}
		\Big\}\le \exp\big(-\nu^2\wt\mD_{ii}^{(\mS_m)}/2\big)
		+ \exp\big\{-\nu^2\wt\mD_{ii}^{(\mS_m)}/(2+2\nu/3)\big\}\\
		& \le 2\exp\big(
		-\nu^2\wt\mD_{ii}^{(\mS_m)}/3\big)
		\end{align*}
		Taking $\nu$ as in (\ref{nu_def}), it can be verified that
		\beq
		\exp\big(
		-\nu^2\wt\mD_{ii}^{(\mS_m)}/3\big)\le 2\exp\{-\log(4(n_{m}+2 l) / \epsilon_m)\} = \epsilon_m/\{2(n_m+2l)\}.\nonumber
		\eeq
		Consequently (\ref{L_C_diff}) holds.

	\noindent
	{\sc 2. Proof of (\ref{L_C_diff2}).}
	
		We bound the second part using the following concentration inequality given by \cite{chung2011spectra}.
	    \bel\label{Ineuality2011}
	    Let $X_{1}, X_{2}, \ldots, X_{m}$ be independent random $N \times N$ Hermitian matrices. Moreover, assume that $\left\|X_{i}-\mathbb{E}\left(X_{i}\right)\right\|_{\max} \leq M$ for all $i,$ and $c^{2}=\left\|\sum\operatorname{var}\left(X_{i}\right)\right\|_{\max} .$ Let $X=\sum X_{i} .$ Then for any $\nu>0$
		$$
		\mathbb{P}(\|X-\mathbb{E}(X)\|_{\max} \geq \nu) \leq 2 N \exp \left(-\frac{\nu^{2}}{2 c^{2}+2 M \nu / 3}\right)
		$$
		\eel
		Denote $E^{i, j} \in \mathbb{R}^{\left(n_{m}+2 l\right) \times\left(n_{m}+2 l\right)}$ with 1 in the $(i, j),(j, i)$ positions and 0 elsewhere, and define
		$
		X^{i, n_{m}+l+j}=(\mathcal{D}^{(\mS_m)}_{i i}\mF^{(\mS_m)}_{j j})^{-1/2}(A^{(\mS_m)}_{i j}-\mathcal{A}^{(\mS_m)}_{i j}) E^{i, n_{m}+l+j}, i=1, \ldots, n_{m}+l, j=1, \ldots, l
		$.
		Then we have
		$$
		\|\wt{C}^{(\mS_m)}-\wt{\mathcal{L}}^{(\mS_m)}\|_{\max}=\Big\|\sum_{i=1}^{n_{m}+l} \sum_{j=1}^{l} X^{i, n_{m}+l+j}\Big\|_{\max},
		$$
due to that $\min_j\mF_{jj}^{(\mS_m)}\ge \min_i \mD_{ii}^{(\mS_m)}$.
As a result, $X^{i, n_{m}+l+j}$ are independent random Hermitian matrices.
We then derive $M$ and $c^2$ in this context and then the results can be obtained by Lemma \ref{Ineuality2011}.
First note that $\mathbb{E}\left[X^{i, n_{m}+l+j}\right]=\zero$. We then have
$$\|X^{i, n_{m}+l+j}\|_{\max} \leq
1/\sqrt{\mD^{(\mS_m)}_{i i}\mathcal{F}^{(\mS_m)}_{j j}} \leq 1/\delta_{m}\defeq M.$$
Next, note that
$		\mathbb{E}\big[\big(X^{i, \ol n_m+j}\big)^{2}\big]=\big(1/\mD^{(\mS_m)}_{i i} \mF^{(\mS_m)}_{j j}\big) \big[\mathcal{A}^{(\mS_m)}_{i j}\big(1-\mathcal{A}^{(\mS_m)}_{i j}\big)\big]$
		$(E^{i i}+E^{\ol n_m+j,\ol n_m+j})\defeq
v_{ij}(E^{i i}+E^{\ol n_m+j,\ol n_m+j})$.
This leads to
		\begin{align*}
		&\Big\|\sum_{i=1}^{n_{m}+l} \sum_{j=1}^{l} \mathbb{E}\left[\left(X^{i, n_{m}+l+j}\right)^{2}\right]\Big\|_{\max}
= \Big\|\sum_{i=1}^{n_{m}+l} \sum_{j=1}^{l}
v_{ij}(E^{i i}+E^{n_{m}+l+j,n_{m}+l+j})
\Big\|_{\max}\\
& =  \max\big\{
\max_{1\le i\le \ol n_m}\sum_j v_{ij}, \max_{1\le j\le l}\sum_i v_{ij}
\big\}\le \frac{1}{\delta_m}\defeq c^2,
		\end{align*}
where the last inequality holds because
\begin{align*}
\sum_j v_{ij} \le \frac{1}{\delta_m}\sum_{j = 1}^{l}\frac{\mathcal{A}^{(\mS_m)}_{i j}}{\mD^{(\mS_m)}_{i i}} = \frac{1}{\delta_m},~~
\sum_i v_{ij} \le \frac{1}{\delta_m}\sum_{i = 1}^{\ol n_m}\frac{\mathcal{A}^{(\mS_m)}_{i j}}{\mF^{(\mS_m)}_{i i}} = \frac{1}{\delta_m}.
\end{align*}
By assumption $\delta_{m}>3 \log \left(n_{m}+2 l\right)+3 \log (4 / \epsilon_m)$, we have $\nu<1 .$ Applying Lemma \ref{Ineuality2011}, we have
		\begin{align}\nonumber
		\mathbb{P}(\Vert \wt{C}^{(\mS_m)}-\wt{\mL}^{(\mS_m)}\Vert_{\max}\ge\nu) &\le 2(n_{m}+2l)\exp\left\{-\frac{2 \log(4(n_{m}+2l)/\epsilon_m)/\delta_{m}}{2/\delta_{m}+2\nu/3\delta_{m}}
		\right\} \nonumber\\
		&\le 2(n_{m}+2l)\exp \big\{-\frac{3\log (4(n_{m}+2l)/\epsilon_m)}{3}\big\} \le \epsilon/2 \nonumber
		\end{align}
This completes the proof.

\end{proof}

{

\bel \label{lemma:master_SBM-eigvec-L2-bound}
Let $\lambda_{1,0}\ge \lambda_{2,0}\ge \cdots \ge \lambda_{K,0}>0$ be the top $K$ singular values of $\mL_0$. Define $\delta_0=\min_i\mD_{0,ii}$. Then for any $\epsilon>0$ and $\delta_0>3l\log(2l)+3\log(4/\epsilon)$, with probability at least $1-\epsilon$ it holds
\begin{equation}\label{equation_master_eigvec_L2_bound}
	\|\wh U_0-U_0Q_0\|_F\le \frac{8\sqrt{6}}{\lambda_{K,0}}\sqrt{\frac{K\log(8l/\epsilon)}{\delta_0}}
\end{equation}
where $Q_0\in \mR^{K\times K}$ is a $K\times K$ orthogonal matrix.
\eel
\begin{proof}
	The proof follows the same procedure as in Theorem \ref{thm_U_diff}.
\end{proof}

\begin{lemma}\label{lemma_master_miscluster}
Define $P_0=\max\limits_{j=1,...,l}(\Theta_0^{\top}\Theta_0)_{j j}$. Denote $\mM$ as the index set of misclustered nodes on the master server. Then for any $\epsilon$ and $\delta_0>3l\log(2l)+3\log(4/\epsilon)$, it holds with probability $1-\epsilon$ that
	\begin{equation}\nonumber
		|\mM|\le \frac{3072P_0K\log(8l/\epsilon)}{\delta_0\lambda_{K,0}^2}
	\end{equation}
	
	\begin{proof}
		Under the procedure in \cite{rohe2011spectral}, it could be verified that \begin{equation}\label{equation_master_misclu_1}
			|\mM|\le8P_0\|\wh U_0-U_0Q_0\|_F^2.
		\end{equation}
		Combining (\ref{equation_master_eigvec_L2_bound}) and (\ref{equation_master_misclu_1}) yields the result.
	\end{proof}
\end{lemma}
}







\scsection{APPENDIX B: Proof of Propositions}
\renewcommand{\theequation}{B.\arabic{equation}}
\setcounter{equation}{0}

\scsubsection{Appendix B.1: Proof of Proposition \ref{prop_U}}\label{Appendix_proof_Prop1}

{\sc Step 1.} We first explore the spectral structure of $\mL$ and $\mL_0$. Construct a matrix $B_{L}\in \mR^{K\times K}$ such that $\mL=\Theta B_{L} \Theta^{\top}$. Define $D_B=\diag (B\Theta ^{\top}\textbf{1}_N)\in \mR^{K\times K}$ where $\textbf{1}_N$ is an $N\times 1$ vector with all entries 1. Denote $\Theta_i$ as the $i$th row of $\Theta$. Note that for any $i, j$,
\begin{equation}
	\mL_{ij}=\frac{\mathcal{A}_{i j}}{\sqrt{\mD_{i i}\mD_{j j}}}=\Theta_iD_{B}^{-1/2}BD_B^{-1/2}\Theta_j^{\top}.
\end{equation}
Consequently, define $B_L=D_B^{-1/2}BD_B^{-1/2}$. It follows $\mL=\Theta B_L \Theta^{\top}$.\\

Similarly, For $\mL_0$, define $D_{B_0}=\diag (B\Theta_0 ^{\top}\textbf{1}_N)\in \mR^{K\times K}$ and $B_{L_0}=D_{B_0}^{-1/2}BD_{B_0}^{-1/2}$, it can be obtained that $\mL_0=\Theta_0 B_{L_0} \Theta_0^{\top}  $.\\

{\sc Step 2.} Denote $\Lambda=\Theta^{\top}\Theta$, $\Lambda_0=\Theta_0^{\top}\Theta_{0}$. Construct $\mL$ and $\mL_0$ as
\begin{align*}
	&\mL=\Theta \Lambda^{-1/2} \Lambda^{1/2} B_L\Lambda^{1/2} \Lambda^{-1/2} \Theta^{\top},\\
	\mL&_0=\Theta_0 \Lambda_0^{-1/2} \Lambda_0^{1/2}B_{L_0} \Lambda_0^{1/2} \Lambda_0^{-1/2}\Theta_0^{\top}.
\end{align*}
Conduct eigen-decompositions as $\Lambda^{1/2} B_L\Lambda^{1/2}=\mu U \mu^{\top}$ and $\Lambda_0^{1/2}B_{L_0} \Lambda_0^{1/2}=\mu_0 U_0 \mu_0^{\top}$, where $\mu$,$\mu_0\in\mR^{K\times K} $ are orthogonal matrices and $U$,$U_0\in \mR^{K\times K}$ are diagonal matrices. By the assumption $m_{0k}/m_k=l/N=r_0$, we have $\Lambda_0=r_0\Lambda$ and $\Theta_0^{\top}\textbf{1}_l=r_0\Theta^{\top}\textbf{1}_N$.  \\

{\sc Step 3.} Recall the eigen-decomposition of $\mL_0$ and $\mL$, by Step 2, we know that $\Lambda^{1/2} B_L\Lambda^{1/2}$ and $\Lambda_0^{1/2}B_{L_0} \Lambda_0^{1/2}$ differ from a scalar multiplication, thus $\mu=\mu_0$.
Subsequently, $\mL$ and $\mL_0$ have the following eigen-decomposition:
\begin{align*}
	&\mL=\Theta \Lambda^{-1/2} \mu U \mu^{\top} \Lambda^{-1/2} \Theta^{\top},\\
	\mL&_0=\Theta_0 \Lambda_0^{-1/2} \mu_0 U_0 \mu_0^{\top} \Lambda_0^{-1/2}\Theta_0^{\top}.
\end{align*}
Further note that $U^{(K)}=\Lambda^{-1/2} \mu$ and $U_0^{(K)}=\Lambda_0^{-1/2} \mu$, then the result naturally holds.

\scsubsection{Appendix B.2: Proof of Proposition \ref{prop_membership} }\label{Appendix_proof_prop2}
\begin{proof}
	We separate the proof into two steps. \\
	In the first step, we show that
	$\mL^{(\mS_m)}$ can be expressed as
	\beq
	\mathcal{L}^{(\mS_{m})}=\Theta^{(\mS_{m})} (\mathcal{D}^{(\mS_{m})}_{B})^{-1 / 2} B (\mathcal{F}^{(\mS_{m})}_{B})^{-1 / 2} \Theta_{0}^{\top}, \label{L_S_m}
	\eeq
	where
	$\mD_B^{(\mS_m)} = \diag\{B\Theta_0^{\top} \mathbf{1}_{l}\}\in \mR^{K\times K}
	$ and
	$\mF_B^{(\mS_m)} = \diag\{B(\Theta^{(\mS_{m})})^{\top} \mathbf{1}_{\ol n_m}\}\in \mR^{K\times K}$.
	In the second step, based on the form in (\ref{L_S_m}), we show that
	$U^{(\mS_m)} = \Theta^{(\mS_{m})}\mu$ is the eigenvector matrix of $\mL^{(\mS_m)}\mL^{(\mS_m)\top}$ and $\mu$ is a full rank matrix.
	This leads to the final result.

	{\sc Step 1.}
	Note that $\bA^{(\mS_m)}\textbf{1}_{l}=\Theta^{(\mS_m)}B\Theta_{0}^{\top}\textbf{1}_{l}$ and $\bA^{(\mS_m)\top}\textbf{1}_{\ol n_m}=\Theta_0B\Theta^{(\mS_m)\top}\textbf{1}_{\ol n_m}$. Therefore, we have $\mD^{(\mS_m)}=\diag\{\bA^{(\mS_m)}\textbf{1}_l\}$ and $\mF^{(\mS_m)}=\diag\{\bA^{(\mS_m)\top}\textbf{1}_{\ol n_m}\}$. Then we have
	\begin{align*}
	& \mD_{ii}^{(\mS_m)}=\Theta_i^{(\mS_{m})}B\Theta_0^{\top}\textbf{1}_{l}=B_{g_i}^{\top}\Theta_{0}^{\top}\textbf{1}_{l}\\
	&\mF_{ii}^{(\mS_m)}=\Theta_{0i}B\Theta^{(\mS_{m})\top}\textbf{1}_{\ol n_m}=B_{g_i}^{\top}\Theta^{(\mS_m)\top}\textbf{1}_{\ol n_m}.
	\end{align*}
	Then it can be obtained that
	\begin{align*}
	\mL_{ij}^{(\mS_m)} &= \frac{{\mathcal{A}}^{(\mS_{m})}_{i j}}{\sqrt{\mathcal{D}^{(\mS_{m})}_{i i} \mathcal{F}^{(\mS_{m})}_{j j}}}
	= \big(B_{g_i}^\top \Theta_0^\top\one_l\big)^{-1/2}\big(\Theta_i^{(\mS_m)\top} B \Theta_{0j}\big)
	\big(B_{g_i}^\top \Theta^{(\mS_m)\top}\one_{\ol n_m}\big)^{-1/2}\\
	& = \e_{g_i}^\top (\mD_B^{(\mS_m)})^{-1/2}
	\big(\Theta_i^{(\mS_m)\top} B \Theta_{0j}\big)
	(\mF_B^{(\mS_m)})^{-1/2}\e_{g_j}\\
	& = \Theta_i^{(\mS_m)\top} (\mD_B^{(\mS_m)})^{-1/2} B (\mF_B^{(\mS_m)})^{-1/2} \Theta_{0j}.
	\end{align*}
	This immediately yields (\ref{L_S_m}).

	{\sc Step 2.}
	In the following we show that the eigen-decomposition of
	$\mathcal{L}^{(\mS_{m})} (\mathcal{L}^{(\mS_{m})})^{\top}$ takes the form
	\beq
	\mathcal{L}^{(\mS_{m})} (\mathcal{L}^{(\mS_{m})})^{\top} = (\Theta^{(\mS_{m})} \mu) \Lambda(\Theta^{(\mS_{m})} \mu)^{\top},\nonumber
	\eeq
	where $U^{(\mS_m)} = \Theta^{(\mS_{m})} \mu\in \mR^{\ol n_m\times K}$ is the eigenvector matrix
	and $\Lambda\in \mR^{K\times K}$ is the diagonal eigenvalue matrix.
	To this end,
	we first write
	$$
	\mathcal{L}^{(\mS_{m})} (\mathcal{L}^{(\mS_{m})})^{\top} =\Theta^{(\mS_{m})} (\mathcal{D}^{(\mS_{m})}_{B})^{-1 / 2} B (\mathcal{F}^{(\mS_{m})}_{B})^{-1 / 2} \Theta_{0}^{\top} \Theta_{0} (\mathcal{F}^{(\mS_{m})}_{B})^{-1 / 2} B (\mathcal{D}^{(\mS_{m})}_{B})^{-1 / 2} (\Theta^{(\mS_{m})})^{\top},
	$$
	$\defeq \Theta^{(\mS_{m})} B_{L} (\Theta^{(\mS_{m})})^{\top}$.
	Define $\Delta = (\Theta^{(\mS_{m})})^{\top} \Theta^{(\mS_{m})}$.
	Then conduct the following eigen-decomposition as $\Delta^{1/2}B_{L}\Delta^{1/2} =
	V\Lambda V^\top$.
	This further implies
	\begin{align*}
	&\Theta^{(\mS_{m})} B_{L} (\Theta^{(\mS_{m})})^{\top}=
	(\Theta^{(\mS_m)}\Delta^{-1/2})\Delta^{1/2}B_{L}\Delta^{1/2}(\Delta^{-1/2}\Theta^{(\mS_m)\top}) \\
	&=
	(\Theta^{(\mS_m)}\Delta^{-1/2})V\Lambda V^\top(\Delta^{-1/2}\Theta^{(\mS_m)\top}) \defeq
	(\Theta^{(\mS_m)} \mu) \Lambda (\Theta^{(\mS_m)} \mu)^\top.
	\end{align*}
	Note that $(\Theta^{(\mS_{m})} \mu)^{\top}(\Theta^{(\mS_{m})} \mu)=I_{K}.$
	By the uniqueness of the eigen-decomposition, we know
	$U^{(\mS_m)} = \Theta^{(\mS_{m})} \mu$
	is the eigenvector matrix of $\mathcal{L}^{(\mS_{m})} (\mathcal{L}^{(\mS_{m})})^{\top} $.
	Further note that the matrix $\mu$ is full rank, then we can conclude that
	$$
	\Theta^{(\mS_{m})}_{i}\mu  = \Theta^{(\mS_{m})}_{j} \mu\Leftrightarrow \Theta^{(\mS_{m})}_{i}=\Theta^{(\mS_{m})}_{j}.
	$$
\end{proof}

\scsubsection{Appendix B.3: Proof of Proposition \ref{prop_diff_worker_spectrum}}\label{Appendix_proof_prop3}
\begin{proof}
	Denote $\bA^{(\mS_m\star)}=\Theta^{(\mS_m)}B\Theta^{(\mS_m)\top}$ and $\mL^{(\mS_m\star)}$ to be its Laplacian matrix, $U^{(\mS_m\star)}$ be the $K$ leading eigenvectors of $\mL^{(\mS_m\star)}$.
	We then have
	\beq
	\|U^{(\mS_{m})}-r_mU_mQ_m\|_F\le \|U^{\mS_m}-U^{(\mS_m\star)}Q_{m1}\|_F +\|U^{(\mS_m\star)}Q_{m1}-r_mU_mQ_{m}\|_F,\nonumber
	\eeq
	where $Q_{m1}$ is another $K\times K$ orthogonal matrix.
	In the following we show that
	\begin{align}
	& \| U^{(\mS_m)}-U^{(\mS_m\star)}Q_{m1}\|_F \le
	\frac{8\sqrt{2} K^2u_0u_m^2\max\{u_0^{1/2},u_m^{1/2}\} \alpha^{(\mS_m)1/2}}{\sigma_{\min}(B)^2b_{\min}^3d_0^2d_m^3},
	\label{spectrum_first_bound}\\
	& \|U^{(\mS_m\star)}Q_{m1}-r_mU_mQ_{m}\|_F \le
	\frac{6\sqrt 2Ku_m\max\{u_0^{1/2},u_m^{1/2}\}\alpha^{(\mS_m)1/2}}{\sigma_{\min}(B)b_{\min}^2d_0d_m^2(d_0+d_m)}+
	\frac{\alpha^{(\mS_m)}}{d_0}\label{spectrum_second_bound}
	\end{align}
	where $Q_{m1}$ is a $K\times K$ orthogonal matrix.
	Then combining (\ref{spectrum_first_bound}) and (\ref{spectrum_second_bound}) yields (\ref{equa_worker_spectrum}).
	The proof is separated into three parts as follows.
	
	\noindent
	{\sc 0. Re-express $\mL^{(\mS_m)}$.}
	
	
	Firstly, we show that
	$\mL^{(\mS_m)}$ can be expressed as
	\beq
	\mathcal{L}^{(\mS_{m})}=\Theta^{(\mS_{m})} (\mathcal{D}^{(\mS_{m})}_{B})^{-1 / 2} B (\mathcal{F}^{(\mS_{m})}_{B})^{-1 / 2} \Theta_{0}^{\top}, \label{L_sm}
	\eeq
	where
	$\mD_B^{(\mS_m)} = \diag\{B\Theta_0^{\top} \mathbf{1}_{l}\}\in \mR^{K\times K}
	$ and
	$\mF_B^{(\mS_m)} = \diag\{B(\Theta^{(\mS_{m})})^{\top} \mathbf{1}_{\ol n_m}\}\in \mR^{K\times K}$.
	Note that $\bA^{(\mS_m)}\textbf{1}_{l}=\Theta^{(\mS_m)}B\Theta_{0}^{\top}\textbf{1}_{l}$ and $\bA^{(\mS_m)\top}\textbf{1}_{\ol n_m}=\Theta_0B\Theta^{(\mS_m)\top}\textbf{1}_{\ol n_m}$. Therefore, we have $\mD^{(\mS_m)}=\diag\{\bA^{(\mS_m)}\textbf{1}_l\}$ and $\mF^{(\mS_m)}=\diag\{\bA^{(\mS_m)\top}\textbf{1}_{\ol n_m}\}$. Then we have
	\begin{align*}
	& \mD_{ii}^{(\mS_m)}=\Theta_i^{(\mS_{m})\top}B\Theta_0^{\top}\textbf{1}_{l}=
	B_{g_i}^{\top}\Theta_{0}^{\top}\textbf{1}_{l}\\
	&\mF_{ii}^{(\mS_m)}=\Theta_{0i}^\top B\Theta^{(\mS_{m})\top}\textbf{1}_{\ol n_m}=B_{g_i}^{\top}\Theta^{(\mS_m)\top}\textbf{1}_{\ol n_m}.
	\end{align*}
	Then it can be obtained that
	\begin{align*}
	\mL_{ij}^{(\mS_m)} &= \frac{\mathcal{A}^{(\mS_{m})}_{i j}}{\sqrt{\mathcal{D}^{(\mS_{m})}_{i i} \mathcal{F}^{(\mS_{m})}_{j j}}}
	= \big(B_{g_i}^\top \Theta_0^\top\one_l\big)^{-1/2}\big(\Theta_i^{(\mS_m)\top} B \Theta_{0j}\big)
	\big(B_{g_i}^\top \Theta^{(\mS_m)\top}\one_{\ol n_m}\big)^{-1/2}\\
	& = \e_{g_i}^\top (\mD_B^{(\mS_m)})^{-1/2}\e_{g_i}
	\big(\Theta_i^{(\mS_m)\top} B \Theta_{0j}\big)
	\e_{g_j}^\top(\mF_B^{(\mS_m)})^{-1/2}\e_{g_j}\\
	& = \Theta_i^{(\mS_m)\top} (\mD_B^{(\mS_m)})^{-1/2} B (\mF_B^{(\mS_m)})^{-1/2} \Theta_{0j},
	\end{align*}
	This immediately yields (\ref{L_sm}).
	Similarly define $\mD_B=\diag\{B\Theta^{\top}\textbf{1}_{N}\}$. We have
	\begin{align}
	\mathcal{L}^{(\mS_{m}\star)}=&\Theta^{(\mS_{m})} (\mathcal{F}^{(\mS_{m})}_{B})^{-1 / 2} B (\mathcal{F}^{(\mS_{m})}_{B})^{-1 / 2} \Theta^{(\mS_m)\top} \label{L_SM_STAR}\\
	\mathcal{L} = &\Theta\mathcal{D}_{B}^{-1 / 2} B \mathcal{D}_{B}^{-1 / 2} \Theta^{\top} \label{L_WHOLE}
	\end{align}
	Now we prove (\ref{spectrum_first_bound}) and (\ref{spectrum_second_bound}) respectively.\\
	
	\noindent
	{\sc 1. Proof of (\ref{spectrum_first_bound}).}
	
	Denote $B^{(\mS_m\star)}=(\Theta^{(\mS_m)\top}\Theta^{(\mS_m)})^{1/2}(\mathcal{F}^{(\mS_{m})}_{B})^{-1 / 2} B (\mathcal{F}^{(\mS_{m})}_{B})^{-1 / 2} (\Theta^{(\mS_m)\top}\Theta^{(\mS_m)})^{1/2}$, $B^{(\mS_m)}=(\Theta^{(\mS_m)\top}\Theta^{(\mS_m)})^{1/2}(\mathcal{D}^{(\mS_{m})}_{B})^{-1 / 2} B $ $(\mathcal{F}^{(\mS_{m})}_{B})^{-1 / 2} (\Theta_{0}^{\top}\Theta_{0})^{1/2}$. It is easy to verify that
	\begin{align*} &\mL^{(\mS_{m}\star)}\mL^{(\mS_{m}\star)\top}=\Theta^{(\mS_{m})}(\Theta^{(\mS_m)\top}\Theta^{(\mS_m)})^{-1/2}B^{(\mS_m\star)}B^{(\mS_m\star)\top}(\Theta^{(\mS_m)\top}\Theta^{(\mS_m)})^{-1/2}\Theta^{(\mS_{m})\top}\\  &\mL^{(\mS_m)}\mL^{(\mS_m)\top}=\Theta^{(\mS_{m})} (\Theta^{(\mS_m)\top}\Theta^{(\mS_m)})^{-1/2}B^{(\mS_m)}B^{(\mS_m)\top}(\Theta^{(\mS_m)\top}\Theta^{(\mS_m)})^{-1/2}\Theta^{(\mS_{m})\top}.
	\end{align*}
	We separate the proof in following three steps.

	\noindent
	{\bf Step 1.1 (Relate $\| U^{(\mS_m)}-U^{(\mS_m\star)}Q_{m1}\|_F$
		to $ \big\|B^{(\mS_{m}\star)}
		B^{(\mS_{m}\star)\top}-B^{(\mS_m)}B^{(\mS_m)\top}\big\|_{\max}$).}

	Denote $\mu^{(\mS_m\star)}, \mu^{(\mS_m)} \in \mathbb{R}^{K\times K}$ as the eigenvectors of $B^{(\mS_{m}\star)}B^{(\mS_{m}\star)\top}$ and $B^{(\mS_{m})}B^{(\mS_{m})\top}$, respectively.
	Then immediately we have
	$U^{(\mS_m)}=\Theta^{(\mS_m)}(\Theta^{(\mS_m)\top}\Theta^{(\mS_m)})^{-1/2}\mu^{(\mS_m)}$, $U^{(\mS_m\star)}=\Theta^{(\mS_m)}(\Theta^{(\mS_m)\top}\Theta^{(\mS_m)})^{-1/2}\mu^{(\mS_m\star)}$.
	Using Lemma 5.1 of \cite{lei2015consistency}, we have
	\begin{align*}
	\|\mu^{(\mS_m)}- \mu^{(\mS_m\star)}Q_{m1}\|_F&\le\frac{2\sqrt{2}K}{\gamma_{m}}\big\|B^{(\mS_{m}\star)}
	B^{(\mS_{m}\star)\top}-B^{(\mS_m)}B^{(\mS_m)\top}\big\|_{\max},
	\end{align*}
	where $\gamma_m$ is the smallest eigenvalue of $B^{(\mS_{m})}B^{(\mS_{m})\top}$.
	Then
	\begin{align*}
	&\|U^{(\mS_m)}-U^{(\mS_m\star)}Q_{m1}\|_F=\|\Theta^{(\mS_m)}(\Theta^{(\mS_m)\top}\Theta^{(\mS_m)})^{-1/2}(\mu^{(\mS_m)}-\mu^{(\mS_m\star)}Q_{m1})\|_{F}\\
	&\le
	\sigma_{\max}\big\{\Theta^{(\mS_m)}(\Theta^{(\mS_m)\top}\Theta^{(\mS_m)})^{-1/2}\big\}
	\|\mu^{(\mS_m)}- \mu^{(\mS_m\star)}Q_{m1}\|_{F}\\
	&\le \frac{2\sqrt{2}K}{\gamma_{m}}\big\|B^{(\mS_{m}\star)}
	B^{(\mS_{m}\star)\top}-B^{(\mS_m)}B^{(\mS_m)\top}\big\|_{\max}
	\end{align*}
	where the last inequality is due to $
	\sigma_{\max}\big\{\Theta^{(\mS_m)}(\Theta^{(\mS_m)\top}\Theta^{(\mS_m)})^{-1/2}\big\} = 1$.
	In Step 1.2 and 1.3 we derive upper bound for
	$ \big\|B^{(\mS_{m}\star)}
	B^{(\mS_{m}\star)\top}-B^{(\mS_m)}B^{(\mS_m)\top}\big\|_{\max}$
	and lower bound for $\gamma_m$ respectively.

	\noindent
	{\bf Step 1.2 (Upper bound for $ \big\|B^{(\mS_{m}\star)}
		B^{(\mS_{m}\star)\top}-B^{(\mS_m)}B^{(\mS_m)\top}\big\|_{\max}$).}
	
	Note here $\Theta^{(\mS_m)\top}\Theta^{(\mS_m)}$ and $\Theta_{0}^{\top}\Theta_{0}$ are diagonal
	matrices.
	Denote $o_i=(\Theta^{(\mS_m)\top}\Theta^{(\mS_m)})^{1/2}_{i i}$ and $p_j=(\Theta_{0}^{\top}\Theta_{0})^{1/2}_{j j}$, $i,j=1,...,K$. Then we have
	\begin{align*}
	B&^{(\mS_{m}\star)}_{ij}=\frac{o_iB_{i j}o_j}{\sqrt{(B_i^{\top}\Theta^{(\mS_m)\top}\textbf{1}_{\ol n_m})(B_{j}^{\top}\Theta^{(\mS_m)\top}\textbf{1}_{\ol n_m}})}\\
	&B^{(\mS_m)}_{i j}=\frac{o_iB_{i j}p_j}{\sqrt{(B_{i}^{\top}\Theta_0^{\top}\textbf{1}_{l})(B_j^{\top}\Theta^{(\mS_m)\top}\textbf{1}_{\ol n_m})}}
	\end{align*}
	For convenience, denote
	$a_i=B_{i}^{\top}\Theta^{(\mS_m)\top}\textbf{1}_{\ol n_m}$,	$b_i=B_{i}^{\top}\Theta_0^{\top}\textbf{1}_{l}$,
	$c_i=l B_{i}^{\top}\Theta^{(\mS_m)\top}\textbf{1}_{\ol n_m}/\ol n_m$,
	$q_i=o_i\sqrt{l}/\sqrt{\ol n_m}$
	then
	\begin{align*}
	\big|\big(B^{(\mS_{m}\star)}&
	B^{(\mS_{m}\star)\top}-B^{(\mS_m)}B^{(\mS_m)\top}\big)_{i j}\big|=\Big|\sum_{k=1}^{K}\Big(\frac{o_io_jB_{ik}B_{jk}o^{2}_k}{\sqrt{a_i a_j}a_k }-\frac{o_io_jB_{ik}B_{jk}p_k^2}{\sqrt{b_{i}b_{j}}a_k }\Big)\Big|\\
	& \le \frac{o_io_j}{a_k}\sum_{k=1}^{K}
	\Big|\frac{o_{k}^{2}}{\sqrt{a_ia_j} }-\frac{p_k^2}{\sqrt{b_ib_j} }\Big|=
	\frac{o_io_j}{a_k}\sum_{k=1}^{K}\Big|\frac{q_{k}^{2}}{\sqrt{c_ic_j} }-\frac{p_k^2}{\sqrt{b_ib_j} }\Big|\\
	&=\frac{o_io_j}{a_k}\sum_{k=1}^{K}\Big|\frac{q_k^{2}\sqrt{b_ib_j}-p_k^2
		\sqrt{c_ic_j}}{\sqrt{c_ic_jb_ib_j}}\Big|\\
	&\le\frac{o_io_j}{a_k}\sum_{k=1}^{K}\Big(\frac{\big|q_k^{2}-p_k^2\big|}{\sqrt{c_ic_j}}+
	\frac{p_k^2|\sqrt{b_ib_j}-\sqrt{c_ic_j}|}{\sqrt{c_ic_jb_ib_j}}\Big)
	\end{align*}
	We then give upper bounds for the two parts respectively as follows.
	First note that  $a_k=B_{k}^{\top}\Theta^{(\mS_m)\top}\textbf{1}_{\ol n_m}\ge Kb_{\min}\ol n_md_m$,
	$c_i=l B_{i}^{\top}\Theta^{(\mS_m)\top}\textbf{1}_{\ol n_m}/\ol n_m\ge l Kb_{\min}d_m\ol n_m/\ol n_m=lKb_{\min}d_m$, and
	$\big|q_k^{2}-p_k^2\big|=\big|\big({l}/{\ol n_m}\Theta^{(\mS_m)\top}\Theta^{(\mS_m)}-\Theta_0^{\top}\Theta_{0}\big)_{k k}\big| \le l\alpha^{(\mS_m)}$.
	This leads to
	\begin{align}
	\frac{o_io_j}{a_k}\sum_k \frac{\big|q_k^{2}-p_k^2\big|}{\sqrt{c_ic_j}}\le
	\frac{u_m\alpha^{(\mS_m)}}{Kb_{\min}^2d_m^2}\label{upper1}
	\end{align}
	Next, for the second part we have
	$b_i=B_{i}^{\top}\Theta_0^{\top}\textbf{1}_{l}\ge Klb_{\min}d_0$, $p_i^2 \le l u_0$.
	Next we discuss the upper bound for $|\sqrt{b_ib_j} - \sqrt{c_ic_j}|$.
	If $b_ib_j\ge c_ic_j$, then we have
	$\sqrt{b_ib_j} - \sqrt{c_ic_j} = \sqrt{(b_i-c_i+c_i)(b_j-c_j+c_j)} - \sqrt{c_ic_j}
	\le \sqrt{c_j|b_i-c_i|}+\sqrt{c_i|b_j-c_j|}+\sqrt{|b_i-c_i||b_j-c_j|}$.
	Otherwise, the upper bound is given by
	$ \sqrt{b_j|b_i - c_i|}+\sqrt{b_i|c_j - b_j|}+ \sqrt{|b_i - c_i||b_j - c_j|}$.
	Consequently we have
	\beq
	|\sqrt{b_ib_j} - \sqrt{c_ic_j}|\le 2\max_i\{\sqrt{b_i}, \sqrt{c_i}\}\max_{i}
	\sqrt{|b_i - c_i|} + \max_{i}
	|b_i - c_j|.\label{bbcc_ine}
	\eeq
	Since $b_i\le Klu_0$, $c_i\le Klu_m$, and
	$|b_i - c_i| = l|B_i^\top (\Theta_0^\top \one_l/l - \Theta^{(\mS_m)\top} \one_{\ol n_m}/\ol n_m)|\le Kl \alpha^{(\mS_m)}$.
	As a result, we have $|\sqrt{b_ib_j} - \sqrt{c_ic_j}|
	\le 2Kl\max\{u_0^{1/2}, u_m^{1/2}\}\alpha^{(\mS_m)1/2} + Kl\alpha^{(\mS_m)}\le
	3Kl \max\{u_0^{1/2}, u_m^{1/2}\}\alpha^{(\mS_m)1/2}$,
	where the inequality is due to that $\alpha^{(\mS_m)}\le \max\{u_0, u_m\}$.
	As a consequence, the upper bound for the second part is
	\beq
	\frac{o_io_j}{a_k}\sum_{k=1}^{K}
	\frac{p_k^2|\sqrt{b_ib_j}-\sqrt{c_ic_j}|}{\sqrt{c_ic_jb_ib_j}}\le \frac{3u_0u_m\max\{u_0^{1/2},u_m^{1/2}\} \alpha^{(\mS_m)1/2}}{K^2b_{\min}^3d_0d_m^2}.\label{upper2}
	\eeq
	Combing the results from (\ref{upper1}) and (\ref{upper2}), we obtain that
	\begin{align*}
	\big\|B^{(\mS_{m}\star)}
	B^{(\mS_{m}\star)\top}-B^{(\mS_m)}B^{(\mS_m)\top}\big\|_{\max}\le
	\frac{4 u_m\max\{u_0^{1/2},u_m^{1/2}\} \alpha^{(\mS_m)1/2}}{Kb_{\min}^3d_0d_m^2}.
	\end{align*}

	\noindent
	{\bf Step 1.3 (Lower bound on $\gamma_m$).}
	Recall that $\gamma_m$ is the smallest eigenvalue of $B^{(\mS_m)}B^{(\mS_m)\top}$.
	Here we have
	$B^{(\mS_m)}B^{(\mS_m)\top}=(\Theta^{(\mS_m)\top}\Theta^{(\mS_m)})^{1/2}(\mathcal{D}^{(\mS_{m})}_{B})^{-1 / 2} B (\mathcal{F}^{(\mS_{m})}_{B})^{-1 / 2} $ $(\Theta_{0}^{\top}\Theta_{0})
	(\mathcal{F}^{(\mS_{m})}_{B})^{-1 / 2} B
	(\mathcal{D}^{(\mS_{m})}_{B})^{-1 / 2}
	(\Theta^{(\mS_m)\top}\Theta^{(\mS_m)})^{1/2}$.
	Specifically $\Theta_0^\top \Theta_0$,
	$\Theta^{(\mS_m)\top}\Theta^{(\mS_m)}$,
	$\mathcal{F}^{(\mS_{m})}_{B}$, and
	$\mathcal{D}^{(\mS_{m})}_{B}$
	are all diagonal matrices.
	As a result, $\lambda_{\min}(\Theta_0^\top\Theta_0)\ge ld_0$,
	$\lambda_{\min}(\Theta^{(\mS_m)\top}\Theta^{(\mS_m)})\ge \ol n_md_m$,
	$\lambda_{\max}(\mathcal{F}^{(\mS_{m})}_{B})\le K\ol n_m u_m$, and
	$\lambda_{\max}(\mathcal{D}^{(\mS_{m})}_{B})\le Klu_0$.
	Therefore we have
	\begin{align*}
	\gamma_m \ge \sigma_{\min}(B)^2 \frac{\ol n_mld_0d_m}{K^2l\ol n_m u_0u_m}
	= \sigma_{\min}(B)^2 \frac{d_0d_m}{K^2 u_0u_m}.
	\end{align*}
	This leads to the final result.

	\noindent
	{\sc 2. Proof of (\ref{spectrum_second_bound}).}\\
	Denote $B^{(\mS_m)}_{L}=(\Theta^{(\mS_m)\top}\Theta^{(\mS_m)})^{1/2}(\mF_B^{(\mS_{m})})^{-1/2}
	B(\mF_B^{(\mS_{m})})^{-1/2}(\Theta^{(\mS_m)\top}\Theta^{(\mS_m)})^{1/2}$ and $B_L=(\Theta^{\top}\Theta)^{1/2}\mD_B^{-1/2}B\mD_B^{-1/2}(\Theta^{\top}\Theta)^{1/2}$.
	According to (\ref{L_SM_STAR}) and (\ref{L_WHOLE}), we have
	\begin{align*}
	&\mL^{(\mS_m\star)}=\Theta^{(\mS_m)}(\Theta^{(\mS_m)\top}\Theta^{(\mS_m)})^{-1/2}B_L^{(\mS_m)}(\Theta^{(\mS_m)\top}\Theta^{(\mS_m)})^{-1/2}\Theta^{(\mS_m)\top}\\
	&\mL=\Theta(\Theta^{\top}\Theta)^{-1/2}B_L(\Theta^{\top}\Theta)^{-1/2}\Theta^{\top}
	\end{align*}
	Denote $f_i=(\Theta^{\top}\Theta)^{1/2}_{i i}$. Note that
	Here we can write
	\begin{align*}
	&(B_L)_{i j}=\frac{f_iB_{i j}f_j}{\sqrt{B_{i}^{\top}\Theta^{\top}\textbf{1}_{N}B_{j}^{\top}
			\Theta^{\top}\textbf{1}_{N}}}\\
	&(B_L^{(\mS_m)})_{i j}=\frac{o_iB_{i j}o_j}{\sqrt{B_{i}^{\top}\Theta^{(\mS_m)\top}\textbf{1}_{\ol n_m}B_{j}^{\top}\Theta^{(\mS_m)\top}\textbf{1}_{\ol n_m}}}
	\end{align*}
	where $f_i = (\Theta^\top \Theta)_{ii}^{1/2}$
	and $B_i$ is the $i$th column of $B$. In the following we prove the upper bound
	in three steps.
	
	\noindent
	{\bf Step 2.1. (Relate $\|U^{(\mS_m\star)}Q_{m1}-r_mU_mQ_{m}\|_F $ to
		$ \|N/\ol n_mB_L - B_L^{(\mS_m)}\|_{\max}$)}
	
	Denote $\xi, \xi^{(\mS_m)}\in \mathbb{R}^{K\times K}$ as the eigenvectors of $B_L$ and $B_L^{(\mS_m)}$, respectively. Assume that the smallest eigenvalue of $B_L^{(\mS_m)}$ is $\tau_{m}$. Note that scalar multiplication does not change the spectrum, using Lemma 5.1 of \cite{lei2015consistency}, we have
	\begin{align*}
	\|\xi^{(\mS_m)}-\xi Q_{m2}\|_F\le\frac{2\sqrt{2}K}{\tau_{m}}\|\frac{N}{\ol n_m}B_L-B_L^{(\mS_m)}\|_{\max}
	\end{align*}
	Note that $U^{(\mS_m\star)}=\Theta^{(\mS_m)}(\Theta^{(\mS_m)\top}\Theta^{(\mS_m)})^{-1/2}\xi
	^{(\mS_m)}$, $U_m=\Theta^{(\mS_m)}(\Theta^{\top}\Theta)^{-1/2}\xi$. Similar to Step 1.1 we have
	\begin{align}\nonumber
	\big\|U&^{(\mS_m\star)}-r_mU_mQ_{m2}\|_F=
	\Big\|\Theta^{(\mS_m)}((\Theta^{(\mS_m)\top}\Theta^{(\mS_m)})^{-1/2}\xi
	^{(\mS_m)}-\sqrt{\frac{N}{\ol n_m}}(\Theta^{\top}\Theta)^{-1/2}\xi Q_{m2})\Big\|_F\\
	&\le \sigma_{\max}\{\Theta^{(\mS_m)}(\Theta^{(\mS_m)\top}\Theta^{(\mS_m)})^{-1/2}\}
	\Big\|\xi
	^{(\mS_m)}-\sqrt{\frac{N}{\ol n_m}}(\Theta^{(\mS_m)\top}\Theta^{(\mS_m)})^{1/2}(\Theta^{\top}\Theta)^{-1/2}\xi Q_{m2}\Big\|_F\nonumber\\
	&\le \Big\|\xi
	^{(\mS_m)}-\sqrt{\frac{N}{\ol n_m}}(\Theta^{(\mS_m)\top}\Theta^{(\mS_m)})^{1/2}(\Theta^{\top}\Theta)^{-1/2}\xi Q_{m2}\Big\|_F,\nonumber
	\end{align}
	where the last inequality is due to that $\sigma_{\max}\{\Theta^{(\mS_m)}(\Theta^{(\mS_m)\top}\Theta^{(\mS_m)})^{-1/2}\}\le 1$.
	Furthermore, it is upper bounded by
	\begin{align}\nonumber
	\Big\|\xi&
	^{(\mS_m)}-\sqrt{\frac{N}{\ol n_m}}(\Theta^{(\mS_m)\top}\Theta^{(\mS_m)})^{1/2}(\Theta^{\top}\Theta)^{-1/2}\xi Q_{m2}\Big\|_F\\
	&\le \|\xi
	^{(\mS_m)}-\xi Q_{m2}\|_F+\|\xi Q_{m2}-{N}^{1/2}/{\ol n_m}^{1/2}(\Theta^{(\mS_m)\top}\Theta^{(\mS_m)})^{1/2}(\Theta^{\top}\Theta)^{-1/2}\xi Q_{m2}\|_F \nonumber\\
	& \le \|\xi
	^{(\mS_m)}-\xi Q_{m2}\|_F+\sigma_{\max}\{I-
	{N}^{1/2}/{\ol n_m}^{1/2}(\Theta^{(\mS_m)\top}\Theta^{(\mS_m)})^{1/2}(\Theta^{\top}\Theta)^{-1/2}\}\|\xi Q_{m2}\|_F\nonumber\\
	& \le  \|\xi
	^{(\mS_m)}-\xi Q_{m2}\|_F+\frac{\alpha^{(\mS_m)}}{d_0}\nonumber
	\end{align}
	where the last inequality holds because
	\begin{align*}
	\Big|1-\sqrt{\frac{No_i^2/\ol n_m }{f_i^2}}\Big|\le\Big|1-\frac{No_i^2/\ol n_m}{f_i^2}\Big|=\frac{\big|f_i^2-No_i^2/\ol n_m\big|}{f_i^2}\le\frac{N\alpha^{(\mS_m)}}{Nd_0}=\frac{\alpha^{(\mS_m)}}{d_0}.
	\end{align*}
	With a simple rotation using $Q_{m1}$, we have
	\beq\nonumber
	\|U^{(\mS_m)\star}Q_{m1}-c_mU_mQ_{m}\|_F \le  \|\xi
	^{(\mS_m)}-\xi Q_{m2}\|_F+\frac{\alpha^{(\mS_m)}}{d_0}
	\eeq
	where $Q_m=Q_{m2}Q_{m1}$.
	
	\noindent
	{\bf Step 2.2. (Upper bound for
		$ \|N/\ol n_mB_L - B_L^{(\mS_m)}\|_{\max}$)}
	
	For convenience , denote $h_i=\ol n_mB_{i}^{\top}\Theta^{\top}\textbf{1}_{N}/N$ and $t_i= \sqrt{\ol n_m}f_i/\sqrt{N}$,
	Then we have
	\begin{align*}
	\Big|(\frac{n_m}{N}B_L-&B_L^{(\mS_m)})_{i j}\Big|=B_{i j}\big|\frac{t_it_j}{\sqrt{h_{i}h_{j}}}-\frac{o_io_j}{\sqrt{a_{i}a_{j}}}\big|\le \big|\frac{t_it_j\sqrt{a_{i}a_j}-o_io_j\sqrt{h_ih_j}}{\sqrt{h_ih_ja_ia_j}}\big|\\
	&\le \frac{o_io_j|\sqrt{a_ia_j}-\sqrt{h_ih_j}|}{\sqrt{h_ih_ja_ia_j}}+
	\frac{|t_it_j-o_io_j|\sqrt{a_ia_j}}{\sqrt{h_ih_ja_ia_j}}
	\end{align*}
	where recall that
	$a_i=B_{i}^{\top}\Theta^{(\mS_m)\top}\textbf{1}_{\ol n_m}$ and $o_i=(\Theta^{(\mS_m)\top}\Theta^{(\mS_m)})^{1/2}_{i i}$.
	We then derive the upper bounds for the above two parts respectively.
	Similar to (\ref{bbcc_ine}), we obtain
	\begin{align*}
	|\sqrt{a_ia_j}-\sqrt{h_ih_j}|&\le  2\max_i\{\sqrt h_i, \sqrt a_i\}\max_{i}\sqrt{|a_i -h_i|} + \max_i|a_i-h_i|\\
	& \le 3K\ol n_m\max\{u_0^{1/2}, u_m^{1/2}\}\alpha^{(\mS_m)1/2},
	\end{align*}
	where the second inequality is due to that
	$h_i\le K\ol n_m u_0$, $a_i\le K\ol n_m u_m$,
	$|a_i-h_i|\le \ol n_m B_i^\top |\Theta^{(\mS_m)\top}\one_{\ol n_m}/\ol n_m -
	\Theta^{\top}\one_{N}/N |\le K\ol n_m\alpha^{(\mS_m)}$
	and $\alpha^{(\mS_m)}\le \max\{u_0, u_m\}$.
	Then we have
	\begin{align}
	\frac{o_io_j|\sqrt{a_ia_j}-\sqrt{h_ih_j}|}{\sqrt{h_ih_ja_ia_j}}\le \frac{3u_m\max\{u_0^{1/2},u_m^{1/2}\}\alpha^{(\mS_m)1/2}}{Kb_{\min}^2d_0d_m}.\label{upper3}
	\end{align}
	Next, note that
	\begin{align*}
	|t_i-o_i|& =\sqrt{\ol n_m}\big|\big[(\frac{\Theta^{\top}\Theta}{N})^{1/2}-(\frac{\Theta^{(\mS_m)\top}\Theta^{(\mS_m)}}{\ol n_m})^{1/2}\big]_{i i}\big|\\
	&=\sqrt{\ol n_m} \frac{|\Theta^{\top}\Theta/N-\Theta^{(\mS_m)\top}\Theta^{(\mS_m)}/\ol n_m|_{i i}}{(\Theta^{\top}\Theta/N)^{1/2}_{i i}+(\Theta^{(\mS_m)\top}\Theta^{(\mS_m)}/\ol n_m)^{1/2}_{i i}} \le \frac{\ol n_m^{1/2}\alpha^{(\mS_m)}}{d_0^{1/2}+d_m^{1/2}}
	\end{align*}
	In addition, we have
	\begin{align*}
	|t_it_j - o_io_j|\le
	2\max_i o_i \max_i |t_i - o_i| + (\max_i|t_i - o_i|)^2
	\le \frac{3\ol n_m \alpha^{(\mS_m)}}{d_0+d_m}
	\end{align*}
	where the last inequality is due to that $o_i\le \ol n_m^{1/2}u_m^{1/2}\le
	\ol n_m^{1/2}$, $\alpha^{(\mS_m)}<1$,
	and $ d_0^{1/2}+d_m^{1/2}\ge d_0+d_m$ with $d_0, d_m\le 1$,  $ (d_0^{1/2}+d_m^{1/2})^2\ge d_0+d_m$.
	As a result, the second part is upper bounded by
	\begin{align}
	\frac{|t_it_j-o_io_j|\sqrt{a_ia_j}}{\sqrt{h_ih_ja_ia_j}}\le \frac{3u_m\alpha^{(\mS_m)}}{Kb_{\min}^2d_0d_m(d_0+d_m)}.\label{upper4}
	\end{align}
	Combining (\ref{upper3}) and (\ref{upper4}), we obtain that
	\begin{align*}
	\Big|\Big(\frac{n_m}{N}B_L-&B_L^{(\mS_m)}\Big)_{i j}\Big|\le \frac{6\max\{u_0^{1/2},u_m^{1/2}\}\alpha^{(\mS_m)1/2}}{Kb_{\min}^2d_0d_m(d_0+d_m)}.
	\end{align*}
	where the inequality holds because $\max\{d_0, d_m\}\le 1/2$ for $K\ge 2$
	and $\alpha^{(\mS_m)}\le \max\{u_0, u_m\}$.

	\noindent
	{\bf Step 2.3 (Lower bound on $\tau_m$)}
	
	Recall that $\tau_m$ is the smallest eigenvalue of $B_L^{(\mS_m)}$.
	Similar to the proof of Step 1.3, we could show
	$
	\tau_m \ge \sigma_{\min}(B){d_m}/({Ku_m})
	$.
	This completes the proof.

\end{proof}

\scsubsection{Appendix B.4: Proof of Proposition \ref{prop_clustering_sufficient_cond}}\label{Appendix_proof_sufficient}
\begin{proof}
Denote $ i_k\in\mC$ as the original index of the $k$th pseudo center, $k=1,...,K$.
	Note that for $l+1\le i\le \ol n_m$ and $k\in$ \{1,...,$K$\} but $k\ne {g_i}$
	\begin{align}
	\|\wh U_{i}^{(\mS_m)}-\wh C_{k}^{(\mS_m)}\|_2&\ge\|\wh C_{g_i}^{(\mS_m)}-\wh C_{k}^{(\mS_m)}\|_2-\|\wh U_{i}^{(\mS_m)}-\wh C_{g_i}^{(\mS_m)}\|_2, \label{bound_igi_ik_hat}\\
	\|\wh C_{g_i}^{(\mS_m)}-\wh C_{k}^{(\mS_m)}\|_2&\ge
	\|U_{i_k}^{(\mS_m)}-U_{i_{g_i}}^{(\mS_m)}\|_2 \nonumber \\
	&-
	\|Q^{(\mS_m)\top}U_{i_k}^{(\mS_m)}-\wh C_{k}^{(\mS_m)}\|_2-
	\|Q^{(\mS_m)\top}U_{i_{g_i}}^{(\mS_m)}-\wh C_{g_i}^{(\mS_m)}\|_2.
	\label{bound_igi_ik}
	\end{align}
	According to Statement D.3 in \cite{rohe2011spectral}
	we have
	\begin{equation}
	\|U_{i_k}^{(\mS_m)}-U_{i_{g_i}}^{(\mS_m)}\|_2\ge\sqrt{\frac{2}{D_m}}\label{Statement3inRohe}
	\end{equation}
	Combining (\ref{Statement3inRohe}) and (\ref{bound_igi_ik}),  we have
	\begin{equation}\nonumber
	\|\wh C_{g_i}^{(\mS_m)}-\wh C_{k}^{(\mS_m)}\|_2\ge\sqrt{\frac{2}{D_m}}-2\zeta_m
	\end{equation}
	Further notice that $P_m=\sqrt{2/D_m}-2\zeta_m$.
	With the condition (\ref{condU}) in Proposition \ref{prop_P_mbiggerthan0},
	using (\ref{bound_igi_ik_hat}), we have
	\begin{align*}
	\|\wh U_{i}^{(\mS_m)}-\wh C_{k}^{(\mS_m)}\|_2&\ge \|\wh C_{g_i}^{(\mS_m)}-\wh C_{k}^{(\mS_m)}\|_2-\|\wh U_{i}^{(\mS_m)}-\wh C_{g_i}^{(\mS_m)}\|_2 \\
	&> P_m-\frac{P_m}{2}=\frac{P_m}{2}> \|\wh U_{i}^{(\mS_m)}-\wh C_{g_i}^{(\mS_m)}\|_2 ,
	\end{align*}
	for any $k\ne g_i$.
	As a result, node $i$ will be correctly clustered.
\end{proof}

\scsubsection{Appendix B.5: Proof of Proposition \ref{prop_P_mbiggerthan0}}\label{Appendix_proof_Pbigger}
\begin{proof}
The final result holds as long as $\zeta_m = o(\ol n_m^{-1/2})$ with probability $1-\epsilon$.
	In the following we prove an upper bound on $\zeta_m$ first. Before that, we clarify the notations of some matrices that will be used in the following proof.
	
	{\sc Notations:}
Denote the centers of clustering after implementing $k$-means on the master server as
	$\wh C\in \mR^{K\times K}$.
Recall that $\mC = \{i_1,\cdots, i_K\}$ collect indexes  of pseudo nodes, where $i_k$ is the index of the node which is closest to the $k$th center.
Correspondingly, let $\wh U_{0c} \defeq (\wh U_{0,i\cdot}:i\in \mC)^\top\in \mR^{K\times K}$ be the mappings of the $K$ pseudo nodes in $\wh U_0$.
In addition, let $U_{0c} =  (U_{0,i\cdot}:i\in\mC)^\top\in\mR^{K\times K}$,
$\wh U_c^{(\mS_m)} \defeq (\wh U_{i\cdot}^{(\mS_m)}, i\in \mC)^\top \in\mR^{K\times K}$,
$U_c^{(\mS_m)}\defeq (U_{i\cdot}^{(\mS_m)}, i\in \mC)^\top \in\mR^{K\times K}$,
and $U_c \defeq (U_{i\cdot}, i\in \mC)^\top \in\mR^{K\times K}$.
Let	$\wh C_u = (\wh C_{\wh g_i}:1\le i\le l)^\top\in\mR^{l\times K}$ collect the $k$-means centers of clusters where each node belongs to.
Next, let $\wh\bg = (\wh g_1,\cdots, \wh g_l)^\top \in\mR^l$.

	Note that $\zeta_m\le \|\wh C^{(\mS_m)}-U_c^{(\mS_m)}\|_F$. We can bound the distance $\|\wh C^{(\mS_m)}-U_c^{(\mS_m)}\|_F$ by using the following inequality:
	\begin{align}
	\|\wh C^{(\mS_m)}-U_c^{(\mS_m)}Q^{(\mS_m)}\|_F&\le \|\wh C^{(\mS_m)}-r_0^{1/2}r_m^{-1/2}U_{0c}Q_mQ^{(\mS_m)}\|_F \label{hat_C_part_1}\\
	&+\|r_0^{1/2}r_m^{-1/2}U_{0c}Q_mQ^{(\mS_m)}-U_c^{(\mS_m)}Q^{(\mS_m)}\|_F. \label{hat_C_part_2}
	\end{align}
	We now bound (\ref{hat_C_part_1}) and (\ref{hat_C_part_2}) respectively.
	
	\noindent
	{\bf Step 1 Upper bound on (\ref{hat_C_part_1})}\\	
	First we have
	\begin{align*}
	\|\wh C^{(\mS_m)}-r_0^{1/2}r_m^{-1/2}U_{0c}Q_mQ^{(\mS_m)}\|_F&=\|r_0^{1/2}r_m^{-1/2}\wh U_{0c}Q_0^{\top}Q_m-r_0^{1/2}r_m^{-1/2}U_cQ_m\|_F\\
	&=r_0^{1/2}r_m^{-1/2}\|\wh U_{0c}Q_0^{\top}-U_{0c}\|_F=r_0^{1/2}r_m^{-1/2}\|\wh U_{0c}-U_{0c}Q_0\|_F,
	\end{align*}
where the equality is due to that $Q_m$, $Q^{(\mS_m)}$ are orthogonal matrices.
	Note that
	\begin{align*}
	\big\|\wh U_{0c}-U_{0c}Q_0\big\|_F\le 2\big\|\wh U_{0c}-\wh C\big\|_F+2\big\|\wh C-U_{0c}Q_0\big\|_F
	\end{align*}
	We bound the two right parts in the following three steps.
	
	\noindent
	{\bf Step 1.1 Upper bound on $\|\wh U_{0c}-\wh C\|_F$.}\\
	Note that rows of $\wh U_{0c}$ are collected as the rows closest to each row in $\wh C$. Then we have
	\begin{align*}
	\|\wh U_{0c}-\wh C\|_F&\le \frac{1}{\sqrt{d_0l}}\|\wh U_0-\wh C_u\|_F\\
	&\le\frac{2}{\sqrt{d_0l}}(\|\wh U_0-U_0Q_0\|_F+\|\wh C_u-U_0Q_0\|_F)
	\end{align*}
	$\|\wh U_0-U_0Q_0\|_F$ has been bounded by Lemma \ref{lemma:master_SBM-eigvec-L2-bound} and $\|\wh C_u-U_0Q_0\|_F$ will be bounded in the next step.
	
	\noindent
	{\bf Step 1.2 Bounds of $\|\wh C-U_cQ_0\|_F$ and $\|\wh C_u-U_0Q_0\|_F$}.\\
	First note that $\wh C$ and $U_c$ are $K$ distinct rows extracted from $\wh C_U$ and $U_0$, respectively.
It suffices to obtain the upper bound for each row of $\wh C$, i.e., $\wh C_j$.
Denote $\wh G_j\subseteq \{1,...,l\}$ as the index sets collecting nodes estimated to be in cluster $j$ and $G_j$ as the index sets collecting nodes truly belonging to cluster $j$, also denote $m_j=|G_j|$ and $\wh m_j=|\wh G_j|$, $j=1,...,K$. Further denote $\wh U_0^{(\wh G_j)}$ as a collection of rows indexed in $\wh G_j$ from $\wh U_0$ and $\wh U_{0i}^{(\wh G_j)}$ as the $i$th row in $\wh U_0^{(\wh G_j)}$.
By the definition, we have
\begin{align}
&\big\|\wh C_j - Q_0^\top C_j\big\|^2 =
\Big\|\frac{\sum_{i=1}^{\wh m_j}\wh U_{0i}^{(\wh G_j)}}{\wh m_j}-\frac{\sum_{i=1}^{\wh m_j}Q_0^\top U_j}{\wh m_j}\Big\|^2
=\frac{1}{\wh m_j^2}\Big\|\sum_{i=1}^{\wh m_j}(\wh U_{0i}^{(\wh G_j)}-Q_0^\top U_j)\Big\|^2\nonumber\\
	&\le \frac{2}{\wh m_j}\sum_{i=1}^{\wh m_j}\|\wh U_{0i}^{(\wh G_j)}-Q_0^\top U_j\|^2
= \frac{2}{\wh m_j}\Big\{\sum_{i\in \wh G_j\cap G_j}\|\wh U_{0i}^{(\wh G_j)}-Q_0^\top U_j\|^2+\sum_{i\in \wh G_j\setminus G_j}\|\wh U_{0i}^{(\wh G_j)}-Q_0^\top U_j\|^2\Big\}\label{delta_C}
\end{align}

According to Lemma \ref{lemma_master_miscluster}, noting that $\delta_0\ge b_{\min}$ and $P_0=lu_0$, an upper bound on the order of $|\wh G_j\setminus G_j|$ and
	a lower bound on $\wh m_j$, comparing to $|G_j|$, can be obtained as
	\begin{equation}
	|\wh G_j\setminus G_j|\le \frac{cu_0K\log(l/\epsilon_l)}{\lambda_{K,0}^2b_{\min}}, \quad \wh m_j\ge ld_0-\frac{cu_0K\log(l/\epsilon_l)}{\lambda_{K,0}^2b_{\min}}\defeq \wt m_j,\label{m_low}
	\end{equation}
	with probability $1-\epsilon_l$,
where $c$ is a finite constant.
By the assumption (C2) and (C3) we can derive that $K\log (l/\epsilon_l)/(\lambda_{K,0}^2 b_{\min})\ll l$.
As a result we have $\wt m_j\ge c_1ld_0$ asymptotically.
	\noindent
	Further note that
	from Lemma \ref{lemma:master_SBM-eigvec-L2-bound}, $\|\wh U_0-U_0Q_0\|_F$ is bounded by
	\begin{equation}\nonumber
	\|\wh U_0-U_0Q_0\|_F\le\frac{8\sqrt{6}}{\lambda_{K,0}}\sqrt{\frac{K\log(8l/\epsilon_l)}{lb_{\min}}}\defeq \wt u_0.
	\end{equation}
since $\delta_0 > lb_{\min}$,
where $\lambda_{0,K}$ is the smallest nonzero singular value of $\mL_0$.
As a result, we have in (\ref{delta_C}) that
\begin{align*}
\sum_{i\in \wh G_j\cap G_j}\|\wh U_{0i}^{(\wh G_j)}-Q_0^\top U_j\|^2+\sum_{i\in \wh G_j\setminus G_j}\|\wh U_{0i}^{(\wh G_j)}-Q_0^\top U_j\|^2\le \wt u_0^2
\end{align*}
with probability at least $1-\epsilon_l$.
Together by using (\ref{m_low}), we have with probability at least $1-\epsilon_l$
	\begin{align*}
	\|\wh C_j-Q_0^\top U_j\|^2 =O(\wt u_0^2/\wt m_j)= o\big(\frac{J_0K\log(l/\epsilon_l) }{ l^2}\big) \quad \text{with}\quad J_0=\frac{1}{b_{\min}\lambda_{K,0}^2},
	\end{align*}
 for $j = 1,\cdots, K$. Subsequently, the bounds can be obtained as
	\begin{align*}
	\|\wh C-U_{0c}Q_0\|_F = o\Big(\frac{KJ_0^{1/2}\log^{1/2}(l/\epsilon_l)}{l}\Big),~~~
	\|\wh C_u-U_0Q_0\|_F = o\Big(\frac{J_0^{1/2}K^{1/2}\log^{1/2} (l/\epsilon_l)}{l^{1/2}}\Big)
	\end{align*}

	\noindent
	{\bf Step 1.3 Bound of (\ref{hat_C_part_1})}\\
	Using the results in Step 1.1 and Step 1.2, considering the number of clusters as a constant and combining the assumptions in Proposition \ref{prop_P_mbiggerthan0}, we have
	\begin{align*}
	 &\|\wh C^{(\mS_m)}-r_0^{1/2}r_m^{1/2}U_{0c}Q_mQ^{(\mS_m)}\|_F \\
&	\le 2 r_0^{1/2}r_m^{-1/2}\big(\frac{1}{\sqrt{d_0l}}\|\wh U_0-U_0Q_0\|_F+\frac{1}{\sqrt{d_0l}}\|\wh C_u-U_0Q_0\|_F+\|\wh C-U_{0c}Q_0\|_F\big)\\
& = o\Big(\frac{r_0^{1/2}r_m^{-1/2}K^{1/2}(\log (l/\epsilon_l))^{1/2}}{ld_0^{1/2}}+
\frac{r_0^{1/2}r_m^{-1/2}K^{1/2}J_0^{1/2}\log (l/\epsilon_l)}{ld_0^{1/2}} +
\frac{r_0^{1/2}r_m^{-1/2}KJ_0^{1/2}\log^{1/2}(l/\epsilon_l)}{l}\Big) \\
	& = o\Big\{\frac{\log^{1/2} (l/\epsilon_l)KJ_0^{1/2}}{l^{1/2}\ol n_m^{1/2} }
\Big\}
	\end{align*}
since $K^2\log(l/\epsilon_l)/(b_{\min}\lambda_{K,0}^2)\ll l$.

	\noindent
	{\bf Step 2: Upper bound on (\ref{hat_C_part_2})}\\
	According to Proposition \ref{prop_diff_worker_spectrum}, we have
	\begin{equation}\nonumber
		\|U^{(\mS_{m})}-r_mU_mQ_m\|_F \le \frac{14\sqrt{2}K^{2}u_m\max\{u_0^{1/2},u_m^{1/2}\}\alpha^{(\mS_m)1/2}}{\sigma_{\min}(B)b^3d_0^2d_m^3(d_0+d_m)}+\frac{\alpha^{(\mS_m)}}{d_0}
\defeq \alpha_m
	\end{equation}
Recall that $U^{(\mS_m)}$ has $K$ distinct rows, which is recorded in $U_c^{(\mS_m)}$.
Then it holds that
	\begin{align*}
	&	\|r_0^{1/2}r_m^{-1/2}U_{0c}Q_mQ^{(\mS_m)}-U_{c}^{(\mS_m)}Q^{(\mS_m)}\|_F
 = \|U_c^{(\mS_m)} - r_0^{1/2}r_m^{-1/2}U_{0c}Q_m\|_F\\
 & = \|U_c^{(\mS_m)} - r_m^{-1/2}U_{c}Q_m\|_F
\le\frac{1}{\sqrt{\ol n_md_m}}\|U^{(\mS_{m})}-r_mU_mQ_m\|_F = o\Big(\frac{\alpha_m}{\ol n_m^{1/2}d_m^{1/2}}\Big).
	\end{align*}

Note that $l \gg K $ and $\alpha^{(\mS_m)} = o(\sigma_{\min}(B)^2/K^4)$ by Condition (C2) and (C3). By the assumptions, we have
and $d_0$, $d_m$, $u_0$, $u_m$ are constants.
It leads to that $\xi_m = o(\ol n_m^{-1/2})$ a.s., which concludes the proof.

\end{proof}

\scsection{APPENDIX C: Proof of Theorems}
\renewcommand{\theequation}{C.\arabic{equation}}
\setcounter{equation}{0}

\scsubsection{Appendix C.1: Proof of Theorem \ref{thm_U_diff}}
\begin{proof}
	Define $\wh H^{(\mS_m)} = 1/\sqrt 2(\wh U^{(\mS_m)\top},
	\wh V^{(\mS_m)\top})^\top\in \mR^{(n_i+2l)\times l}$.
	In addition, let $H^{(\mS_m)} = 1/\sqrt 2(U^{(\mS_m)\top},
	V^{(\mS_m)\top})^\top\in \mR^{(n_i+2l)\times l}$ be its population version.
	By Lemma 5.1 of \cite{lei2015consistency}
	we have
	\beq
	\big\|\wh H^{(\mS_m)}\wh H^{(\mS_m)\top}-H^{(\mS_m)}H^{(\mS_m)\top}
	\big\|_F \le \frac{2K}{\lambda_{K,m}}\big\|\wt L^{(\mS_m)}-\wt \mL^{(\mS_m)}\big\|_{\max}.\label{HH_bound}
	\eeq
	Note that
	\begin{align}\nonumber
	&\wh H^{(\mS_m)}\wh H^{(\mS_m)\top}-H^{(\mS_m)}H^{(\mS_m)\top}\\
	& = \begin{pmatrix}
	\frac{1}{2} \hat{U}^{(\mS_m)} (\hat{U}^{(\mS_m)})^{\top}-\frac{1}{2} U^{(\mS_m)} (U^{(\mS_m)})^{\top} & \frac{1}{2} \hat{U}^{(\mS_m)} (\hat{V}^{(\mS_m)})^{\top}-\frac{1}{2} U^{(\mS_m)} (V^{(\mS_m)})^{\top}\\
	\frac{1}{2} \hat{V}^{(\mS_m)} \hat{U}^{(\mS_m)}{\top}-\frac{1}{2} V^{(\mS_m)} (U^{(\mS_m)})^{\top} & \frac{1}{2} \hat{V}^{(\mS_m)} (\hat{V}^{(\mS_m)})^{\top}-\frac{1}{2} V^{(\mS_m)}(V^{(\mS_m)})^{\top}
	\end{pmatrix} \nonumber
	\end{align}
	This implies
	$
	\big\|\wh H^{(\mS_m)}\wh H^{(\mS_m)\top}-
	H^{(\mS_m)}H^{(\mS_m)\top}\big\|_F\ge
	{1}/{2}\big\|\wh U^{(\mS_m)}\wh U^{(\mS_m)\top}-U^{(\mS_m)}U^{(\mS_m)\top}\big\|_F
	\ge
	{1}/{2}\big\|\wh U^{(\mS_m)}-U^{(\mS_m)}Q^{(\mS_m)}\big\|_F
	$.
	Then the result can be immediately obtained by using (\ref{HH_bound})
	and Proposition \ref{prop_L_bound}.
\end{proof}

\scsubsection{Appendix C.2: Proof of Theorem \ref{thm_misclustered_nodes}}
\begin{proof}
{Denote $E_m$ as the index sets where nodes are misclustered on server $m$ and let $e^{(\mS_m)} = |E_m|$. Using Proposition \ref{prop_clustering_sufficient_cond} and Proposition \ref{prop_P_mbiggerthan0}, $e^{(\mS_m)}$ can be upper bounded with probability $1 - \epsilon_l$ by
\begin{align*}
e^{(\mS_m)}&=\sum\limits_{i\in E_m}1\le\frac{4\ol n_m}{c^2}\sum\limits_{i\in E_m}\big\|\wh U_{i}^{(\mS_m)}-\wh C_{g_i}^{(\mS_m)}\big\|_2^2,
\end{align*}
where $c$ is a constant.
 Note that we have we have $ \|\wh U_{i}^{(\mS_m)}-\wh C_{g_i}^{(\mS_m)}\|_2\le
	\| \wh U_{i}^{(\mS_m)}- Q^{(\mS_m)\top}U^{(\mS_m)}_{i}\|_2+
	\|Q^{(\mS_m)\top}U^{(\mS_m)}_{i}- C^{(\mS_m)}_{g_i}\|_2+
	\| C^{(\mS_m)}_{g_i}- \wh C^{(\mS_m)}_{g_i}\|_2$,
	where $Q^{(\mS_m)}$ is defined in Theorem \ref{thm_U_diff}.
	This yields
	\begin{align}\nonumber
	 e^{(\mS_m)}
	 &\le \frac{12\ol n_m}{c^2}\sum\limits_{i\in E_m}\big(\| \wh U_{i}^{(\mS_m)}- Q^{(\mS_m)\top}U^{(\mS_m)}_{i}\|_2^2+
	\|Q^{(\mS_m)\top}U^{(\mS_m)}_{i}- C^{(\mS_m)}_{g_i}\|_2^2\nonumber\\
	&\qquad \qquad+
	\| C^{(\mS_m)}_{g_i}- \wh C^{(\mS_m)}_{g_i}\|_2^2
	\big)\nonumber\\
	&\le \frac{12\ol n_m}{c^2}\big( \big\|\wh U^{(\mS_m)} - Q^{(\mS_m)\top}U^{(\mS_m)}\big\|_F^2+\|U^{(\mS_m)}-r_m^{-1/2}U_mQ_m\|_F^2\nonumber\\
	&\qquad\qquad+lu_m\|\wh C^{(\mS_m)}-C^{(\mS_m)}\|_F^2\big) \label{equation_final_three_bound}
	.
	\end{align}
Note that $n_mu_m\|\wh C^{(\mS_m)}-C^{(\mS_m)}\|_F^2=u_ml\|\wh C-U_{0c}Q_0\|_F^2$
by the proof procedure in Appendix \ref{prop_P_mbiggerthan0}.
Further combining the results from Theorem \ref{thm_U_diff}, Proposition \ref{prop_diff_worker_spectrum} and the proof of Proposition \ref{prop_P_mbiggerthan0}, each of which bounds one of the three parts in (\ref{equation_final_three_bound}), based on the assumptions, we have
\begin{align*}
\cR^{(\mS_m)} &= \frac{e^{(\mS_m)}}{\ol n_m} \le o\left(\frac{K^2\log(l/\epsilon_l)}{b_{\min}l\lambda_{K,0}^2}+\frac{K\log(4(n_m+2l)/\epsilon_m)}{\lambda_{K,m}\delta_m}+\frac{K^4\alpha^{(\mS_m)}}{\sigma_{\min}(B)^2b_{\min}^6}\right)
\end{align*}
with probability at least $1-\epsilon_m-\epsilon_l$.}

\end{proof}

\end{appendices}
\end{document}